\documentclass[12pt]{article}
\usepackage{amssymb}
\usepackage{graphicx}
\begin{document}
\newcommand{\beq}{\begin{equation}}
\newcommand{\eeq}{\end{equation}}
\newcommand{\beqa}{\begin{eqnarray}}
\newcommand{\eeqa}{\end{eqnarray}}
\newcommand{\beqar}{\begin{eqnarray*}}
\newcommand{\eeqar}{\end{eqnarray*}}
\newcommand{\al}{\alpha}
\newcommand{\be}{\beta}
\newcommand{\del}{\delta}
\newcommand{\D}{\Delta}
\newcommand{\eps}{\epsilon}
\newcommand{\ga}{\gamma}
\newcommand{\Ga}{\Gamma}
\newcommand{\ka}{\kappa}
\newcommand{\nn}{\nonumber}
\newcommand{\inn}{\!\cdot\!}
\newcommand{\h}{\eta}
\newcommand{\ii}{\iota}
\newcommand{\kk}{\varphi}
\newcommand\F{{}_3F_2}
\newcommand{\la}{\lambda}
\newcommand{\La}{\Lambda}
\newcommand{\na}{\prt}
\newcommand{\Om}{\Omega}
\newcommand{\om}{\omega}
\newcommand{\p}{\Phi}
\newcommand{\sig}{\sigma}
\renewcommand{\t}{\theta}
\newcommand{\z}{\zeta}
\newcommand{\ssc}{\scriptscriptstyle}
\newcommand{\eg}{{\it e.g.,}\ }
\newcommand{\ie}{{\it i.e.,}\ }
\newcommand{\labell}[1]{\label{#1}} 
\newcommand{\reef}[1]{(\ref{#1})}
\newcommand\prt{\partial}
\newcommand\veps{\varepsilon}
\newcommand{\pol}{\varepsilon}
\newcommand\vp{\varphi}
\newcommand\ls{\ell_s}
\newcommand\cF{{\cal F}}
\newcommand\cA{{\cal A}}
\newcommand\cS{{\cal S}}
\newcommand\cT{{\cal T}}
\newcommand\cV{{\cal V}}
\newcommand\cL{{\cal L}}
\newcommand\cM{{\cal M}}
\newcommand\cN{{\cal N}}
\newcommand\cG{{\cal G}}
\newcommand\cK{{\cal K}}
\newcommand\cH{{\cal H}}
\newcommand\cI{{\cal I}}
\newcommand\cJ{{\cal J}}
\newcommand\cl{{\iota}}
\newcommand\cP{{\cal P}}
\newcommand\cQ{{\cal Q}}
\newcommand\cg{{\tilde {{\cal G}}}}
\newcommand\cR{{\cal R}}
\newcommand\cB{{\cal B}}
\newcommand\cO{{\cal O}}
\newcommand\tcO{{\tilde {{\cal O}}}}
\newcommand\bz{\bar{z}}
\newcommand\bb{\bar{b}}
\newcommand\ba{\bar{a}}
\newcommand\bg{\bar{g}}
\newcommand\bc{\bar{c}}
\newcommand\bw{\bar{w}}
\newcommand\bX{\bar{X}}
\newcommand\bK{\bar{K}}
\newcommand\bA{\bar{A}}
\newcommand\bH{\bar{H}}
\newcommand\bF{\bar{F}}
\newcommand\bxi{\bar{\xi}}
\newcommand\bphi{\bar{\phi}}
\newcommand\bpsi{\bar{\psi}}
\newcommand\bprt{\bar{\prt}}
\newcommand\bet{\bar{\eta}}
\newcommand\btau{\bar{\tau}}
\newcommand\hF{\hat{F}}
\newcommand\hA{\hat{A}}
\newcommand\hT{\hat{T}}
\newcommand\htau{\hat{\tau}}
\newcommand\hD{\hat{D}}
\newcommand\hf{\hat{f}}
\newcommand\hK{\hat{K}}
\newcommand\hg{\hat{g}}
\newcommand\hp{\hat{\Phi}}
\newcommand\hi{\hat{i}}
\newcommand\ha{\hat{a}}
\newcommand\hb{\hat{b}}
\newcommand\hQ{\hat{Q}}
\newcommand\hP{\hat{\Phi}}
\newcommand\hS{\hat{S}}
\newcommand\hX{\hat{X}}
\newcommand\tL{\tilde{\cal L}}
\newcommand\hL{\hat{\cal L}}
\newcommand\tG{{\tilde G}}
\newcommand\tg{{\tilde g}}
\newcommand\tphi{{\widetilde \Phi}}
\newcommand\tPhi{{\widetilde \Phi}}
\newcommand\te{{\tilde e}}
\newcommand\tk{{\tilde k}}
\newcommand\tf{{\tilde f}}
\newcommand\tH{{\tilde H}}
\newcommand\ta{{\tilde a}}
\newcommand\tb{{\tilde b}}
\newcommand\tc{{\tilde c}}
\newcommand\td{{\tilde d}}
\newcommand\tm{{\tilde m}}
\newcommand\tmu{{\tilde \mu}}
\newcommand\tnu{{\tilde \nu}}
\newcommand\talpha{{\tilde \alpha}}
\newcommand\tbeta{{\tilde \beta}}
\newcommand\trho{{\tilde \rho}}
 \newcommand\tR{{\tilde R}}
\newcommand\teta{{\tilde \eta}}
\newcommand\tF{{\widetilde F}}
\newcommand\tK{{\tilde K}}
\newcommand\tE{{\widetilde E}}
\newcommand\tpsi{{\tilde \psi}}
\newcommand\tX{{\widetilde X}}
\newcommand\tD{{\widetilde D}}
\newcommand\tO{{\widetilde O}}
\newcommand\tS{{\tilde S}}
\newcommand\tB{{\tilde B}}
\newcommand\tA{{\widetilde A}}
\newcommand\tT{{\widetilde T}}
\newcommand\tC{{\widetilde C}}
\newcommand\tV{{\widetilde V}}
\newcommand\thF{{\widetilde {\hat {F}}}}
\newcommand\Tr{{\rm Tr}}
\newcommand\tr{{\rm tr}}
\newcommand\STr{{\rm STr}}
\newcommand\hR{\hat{R}}
\newcommand\M[2]{M^{#1}{}_{#2}}
\newcommand\MZ{\mathbb{Z}}
\newcommand\MR{\mathbb{R}}
\newcommand\bS{\textbf{ S}}
\newcommand\bI{\textbf{ I}}
\newcommand\bJ{\textbf{ J}}

\begin{titlepage}
\begin{center}

\vskip 0.5 cm
{\LARGE \bf 
The emergence of inherently 9-dimensional \\  \vskip 0.25 cm one-loop effective action from T-duality
} \\
\vskip 1.25 cm
 Mohammad R. Garousi \footnote{garousi@um.ac.ir}

\vskip 1 cm
{{\it Department of Physics, Faculty of Science, Ferdowsi University of Mashhad\\}{\it P.O. Box 1436, Mashhad, Iran}\\}
\vskip .1 cm
 \end{center}

\begin{abstract}
Recent studies suggest that applying the Buscher rules to the dimensional reduction of ten-dimensional, one-loop effective actions  generate "purely stringy" couplings in nine dimensions that cannot be lifted to a local, covariant form in ten dimensions. We investigate this phenomenon at order $\alpha'^3$ in type IIA string theory. By computing the circular reduction of the one-loop Chern-Simons term and pure-gravity couplings in type IIA theory and applying the T-duality transformation to the resulting couplings, we derive their counterparts in the type IIB effective action. We demonstrate that the resulting nine-dimensional type IIB couplings are invariant under S-duality without requiring contributions from the tree-level effective action or non-perturbative effects.  As a consistency check, we show that the nine-dimensional type IIB couplings, when reduced on a K3 surface, reproduce the known heterotic string couplings on \( T^5 \) at order \( \alpha' \), via the duality between the two theories.

\end{abstract}

\end{titlepage}

\section{Introduction}

    The  spacetime effective action in string theory at the critical dimension 10 is characterized by a double expansion in the world-sheet genus $g$ and the string scale $\alpha'$, which governs the derivative expansion. A range of complementary techniques—including the S-matrix method \cite{Schwarz:1982jn,Gross:1986iv,Gross:1986mw}, the non-linear sigma model \cite{Grisaru:1986kw,Grisaru:1986vi,Tseytlin:1988rr}, supersymmetry \cite{Ozkan:2024euj,Green:1998by}, T-duality \cite{Giveon:1994fu,Alvarez:1994dn,Garousi:2017fbe},  S-duality \cite{Schwarz:1996bh,Green:1997tv}, and U-duality \cite{Green:2010wi}—are employed to determine the coefficients of this expansion. The dualities impose distinct constraints: T-duality relates effective actions at different orders in $\alpha'$ while preserving the genus, whereas S-duality operates more profoundly, connecting 10-dimensional couplings  across different genera and introducing non-perturbative effects. A striking application of this principle is that imposing S-duality on the classical type IIB effective action at order $\alpha'^3$ uniquely determines the structure of its higher-genus and non-perturbative completions \cite{Green:1997tv,Green:1997di}.
    
To impose the T-duality constraint on the effective action, it is necessary to compactify at least one spatial dimension on a circle. If the circle's radius is treated as a dimensionless parameter, the resulting nine-dimensional theory acquires an additional expansion parameter alongside the $\alpha'$- and $g$-expansions: an expansion in the circle's radius \cite{Garousi:2025xdz}.
At the classical level, the effective action is background-independent \cite{Garousi:2022ovo}. The nine-dimensional effective action can therefore be derived from the Kaluza-Klein (KK) reduction of the ten-dimensional theory. Consequently, the radius-dependence of the classical couplings is known exactly from this KK reduction. This property, in fact, allows one to determine the classical effective action by imposing T-duality on its KK reduction \cite{Garousi:2023kxw,Garousi:2020gio,Wulff:2024ips,Ameri:2025bei}.
At the loop level, however, the effective action is background-dependent \cite{Garousi:2025xdz}. The nine-dimensional effective action is then distinctly different from its ten-dimensional counterpart, and the radius-dependence of the couplings generally cannot be derived from the KK reduction of the ten-dimensional terms. In this case, the couplings must be expressed as a radius expansion.

The leading large-radius term of this expansion is obtained from the KK reduction of the ten-dimensional couplings. In contrast, the leading small-radius terms follow by applying a T-duality transformation to this large-radius result. These small-radius terms—along with all other possible terms in the radius expansion—represent inherently nine-dimensional couplings that cannot be derived from the KK reduction of the ten-dimensional theory \cite{Garousi:2025xdz}. They may instead be determined through explicit S-matrix calculations incorporating the effects of compactification on a circle. This compactification replaces continuous momentum in one direction with discrete KK momentum, while also introducing winding modes along the circle  \cite{Green:1982sw}. While the KK momentum contribution to  the nine-dimensional, loop-level effective action originates from the KK reduction of ten-dimensional couplings, the winding mode contribution cannot be found this way. Ultimately, the full radius expansion must be constructed to ensure the nine-dimensional, loop-level effective action is invariant under T-duality \cite{Garousi:2025xdz}.

In fact,   the four-graviton nine-dimensional  one-loop amplitude in type II string theory has been calculated from eleven-dimensional supergravity compactified on \(T^2\) in \cite{Green:1997as}. It was shown that, at eight-derivative order, the moduli dependence of the nine-dimensional \(R^4\) couplings consists of two distinct parts: one remains non-zero  when the volume of the torus shrinks to zero, while the other vanishes in this limit. The former corresponds to nine-dimensional type IIB couplings that arise as the KK reduction of the ten-dimensional type IIB \(R^4\) couplings, whereas the latter are nine-dimensional type IIB couplings that cannot originate from the ten-dimensional \(R^4\) couplings  \cite{Green:1997as}.

By applying T-duality to the circularly reduced ten-dimensional one-loop effective action of type IIA theory, we derive the corresponding nine-dimensional effective action for type IIB theory at order \(\alpha'^3\) in the small-radius limit. We show that, while S-duality in the large-radius limit requires the inclusion of both tree-level and non-perturbative effects \cite{Green:1997tv,Green:1997di,Liu:2019ses}, the leading small-radius couplings we obtain are individually S-duality invariant. This invariance is explicitly verified by performing the dimensional reduction and T-duality transformation, and then demonstrating that the resulting nine-dimensional type IIB couplings are S-duality invariant in the absence of Ramond-Ramond (RR) fields. These symmetries also appear in the nine-dimensional \(R^4\) couplings reported in \cite{Green:1997as}.

A key observation is that S-duality forces the radius expansion of the type II one-loop effective action at order \(\alpha'^3\) to truncate, leaving only the large- and small-radius limits \cite{Garousi:2025xdz}. The argument follows from the known S-duality properties of the nine-dimensional type IIB one-loop couplings at this order. While these couplings require combination with tree-level and non-perturbative terms to be invariant in the large-radius limit \cite{Green:1997tv,Green:1997di,Liu:2019ses}, we demonstrate that they are inherently invariant in the small-radius limit. Given that the tree-level and non-perturbative sectors are already accounted for, and that there are no two-loop or higher corrections to couplings at this derivative order \cite{Green:1998by,Berkovits:1997pj,Basu:2011he,Bossard:2014lra,DHoker:2005vch}, any additional one-loop terms in a full radius expansion would, by S-duality, need to be separately invariant. The only consistent possibility is for such terms to be absent, thereby confirming the predicted truncation.

 As a further check on the nine-dimensional type IIB couplings at order \(\alpha'^3\), we use the duality between type IIB theory on \(S^1 \times K3\) and heterotic theory on \(T^5\) \cite{Hull:1994ys,Aspinwall:1996mn}. We dimensionally reduce the couplings on a K3 manifold and demonstrate that, while their large-radius limit yields no terms at order \(\alpha'\), their small-radius limit produces five-dimensional couplings at order \(\alpha'\). These are shown to be consistent with the expected couplings of the heterotic theory compactified on \(T^5\), following an appropriate field redefinition that maps the type IIB radius parameter to the heterotic dilaton.
 Given that the dilaton dependence in the heterotic theory at order $\alpha'$—specifically, the factor of \( e^{-2\Phi} \)—is exact \cite{Ellis:1987dc,Ellis:1989fi}, this correspondence implies that the radius dependence of the type IIB couplings is also exact. This result reinforces the conclusion that the nine-dimensional couplings exist in only two distinct forms: one for the large-radius limit and one for the small-radius limit, with no interpolating corrections.

    This paper is structured as follows. In Section 2, we detail the KK reduction of the type IIA Chern-Simons term at order $\alpha'^3$, perform its T-duality transformation to derive the corresponding type IIB couplings, and demonstrate the S-duality invariance of the result. Subsection 2.5 further shows that the K3 reduction of these nine-dimensional, small-radius type IIB couplings yields four-derivative couplings consistent with the heterotic theory on $T^5$. Section 3 extends this analysis to the pure gravity couplings of type IIA, examining their KK reduction, transformation under T-duality and S-duality, and the equivalence of their K3-reduced type IIB form with heterotic theory on $T^5$. Our conclusions are presented in Section 4. All computations were performed using the ``xAct'' package \cite{Nutma:2013zea}.

\section{Chern-Simons term at order $\alpha'^3$}

    The well-known duality between M-theory on a circle and type IIA string theory is reflected in their effective actions: the KK reduction of 11-dimensional supergravity yields 10-dimensional type IIA supergravity \cite{Huq:1983im,Becker:2007zj}, and M-theory's eight-derivative corrections produce one-loop corrections in type IIA string theory. However, the complete eight-derivative couplings for M-theory's massless bosonic fields—the metric and the three-form—are not yet known. In this paper, we focus on the subset of these couplings that have been determined: namely, the Chern-Simons term \cite{Green:1984sg,Duff:1995wd,Peeters:2000qj,Hyakutake:2006aq} where the three-form appears linearly and  the pure gravity terms \cite{Vafa:1995fj,Green:1997di,Russo:1997mk,Peeters:2000qj}.

\subsection{10D Chern-Simons term in type IIA}

    The KK reduction of the eight-derivative Chern-Simons term in M-theory produces the corresponding Chern-Simons term in type IIA theory \cite{Duff:1995wd}, as well as some gauge-invariant couplings involving RR fields \cite{Garousi:2025dil}. The latter couplings are not discussed here, as our focus is on NS-NS couplings. The Chern-Simons term in type IIA theory is 
\beqa
\bS^{CS}_{\rm IIA}&\!\!\!\!\!=\!\!\!\!\!&-\frac{2}{\kappa^2}\frac{\pi^2\alpha'^3}{2^{11}.3^2}\int d^{10}x\sqrt{-G} \,\epsilon_{10}^{\alpha \beta \gamma \mu \nu \kappa 
\lambda \theta \delta \sigma }B_{\alpha\beta}\Big[3  R_{\gamma \mu 
}{}^{\eta \epsilon } R_{\lambda \theta \rho \omega } 
R_{\nu \kappa \eta \epsilon } R_{\delta\sigma 
 }{}^{\rho \omega } \nn\\&&\qquad\qquad\qquad\qquad\qquad\qquad\qquad\qquad\quad- 12  
R_{\gamma \mu }{}^{\eta \epsilon } R_{\lambda 
\theta \omega \epsilon } R_{\nu \kappa \rho \eta } R_{ \delta\sigma }{}^{\rho \omega }\Big]\!,\labell{reduce0}
\eeqa
where $\epsilon_{10}$ denotes the Levi-Civita tensor in ten dimensions, and $\kappa^2=\frac{1}{\pi}(2\pi\sqrt{\alpha'})^8$. The absence of a dilaton in the above action identifies it as the one-loop effective action. Since 11-dimensional M-theory lacks a fundamental string, the KK reduction of its effective action correctly produces the 10-dimensional type IIA couplings. In other words, the above action is valid for any radius of the 11th direction $R_{11}=g_s\sqrt{\alpha'}$. However, this is not the case for the KK reduction of type IIA theory, whose fundamental object is a string originating from an M2-brane wrapped on the circle of the 11th dimension. Consequently, the lower-dimensional one-loop effective action of type IIA string theory may receive both winding and KK contributions. The KK contribution, which we calculate in the following section, is valid only in the large-radius limit.\footnote{       The work in \cite{Liu:2013dna} examined the KK reduction incorporating the B-field via a connection with torsion and analyzed its T-duality transformations in the zero-radion case.}

\subsection{9D Chern-Simons term in type IIA }

The nine-dimensional one-loop couplings in type IIA theory for the large-radius limit can be obtained from the KK reduction of the ten-dimensional one-loop couplings. When one spatial dimension is compactified on a circle of dimensionless radius  \( R=e^{\vp/2 }\), the NS-NS fields reduce according to \cite{Maharana:1992my, Kaloper:1997ux} as:
\beqa  
G_{\mu\nu} = \left(\matrix{\bg_{ab} + e^{\varphi} g_{a} g_{b} &  e^{\varphi} g_{a} \cr e^{\varphi} g_{b} & e^{\varphi} &}\!\!\!\!\!\right), \quad  
B_{\mu\nu} = \left(\matrix{\bb_{ab} + b_{[a} g_{b]} & b_{a} \cr -b_{b} & 0 &}\!\!\!\!\!\right), \quad  
\Phi = \bar{\phi} + \varphi/4,  
\labell{reduc}  
\eeqa 
where indices \(a,b\) denote directions orthogonal to the Killing coordinate \(y\). The resulting nine-dimensional base space fields are: \(\bg_{ab}\) (metric), \(\bb_{ab}\) (antisymmetric tensor), \(\bar{\phi}\) (dilaton), \(\varphi\) (radion), and the vector fields \(g_a\) and \(b_a\).

The reduction of the Levi-Civita symbol $\epsilon'_{10}=\sqrt{-G} \,\epsilon_{10}$ is trivial, as
\beqa
\sqrt{-G} \,\epsilon_{10}^{y abcdefghi }=\sqrt{-\bg} \,\epsilon_{9}^{abcdefghi},
\eeqa
where $\epsilon_{9}$ denotes the Levi-Civita tensor in nine dimensions.      The reduction of the Riemann curvature tensor can also be calculated using the reduction of the metric in \reef{reduc}, and the integral over the Killing coordinate $y$ becomes $2\pi$. A similar calculation was performed in \cite{Garousi:2025dil} when reducing the M-theory Chern-Simons term to produce the type IIA coupling \reef{reduce0}. This reduction produces the nine-dimensional Chern-Simons term \cite{Antoniadis:1997eg}, as well as other terms that can be rendered gauge-invariant after the inclusion of specific non-gauge-invariant total derivative terms. By following the method of \cite{Garousi:2025dil}, one obtains:
\beqa
S^{CS}_{\rm IIA}&\!\!\!=\!\!\!&-\frac{2}{\kappa^2}\frac{\pi^3\alpha'^3}{2^{10}.3^2}\int d^{9}x\sqrt{-\bg} \,\epsilon_{9}^{abcdefghi}\Big[b_a\cL_{ bcdefghi}\!+\!W_{hi}\cL^W_{ abcdefg }\!+\!\bar{H}_{ghi}\cL^{\bar{H}}_{abcdef}\Big]\!,\labell{reduce1}
\eeqa
which includes the two standard terms of the nine-dimensional Chern-Simons form:
\beqa
\cL_{ bcdefghi}&=&24 R_{bc}{}^{jk} R_{dej}{}^{l} 
R_{fgk}{}^{m} R_{hilm}- 6 
R_{bc}{}^{jk} R_{dejk} R_{fg}{}^{lm} 
R_{hilm} .
\eeqa
The Lagrangian $\cL^W_{abcdefg}$ comprises 11 terms. They are
\beqa
\cL^W_{abcdefg }&=&24 e^{2 \vp} R_{fgkl} V_{b}{}^{j} V_{c}{}^{k} V_{de} 
V_{j}{}^{l} \nabla_{a}\vp + 12 e^{\vp} 
R_{de}{}^{kl} R_{fgkl} V_{a}{}^{j} 
\nabla_{c}V_{bj} - 18 e^{2 \vp} R_{fgkl} V_{a}{}^{j} 
V_{b}{}^{k} V_{c}{}^{l} \nabla_{e}V_{dj}\nn\\&& - 6 e^{2 \vp} 
R_{fgkl} V_{a}{}^{j} V_{bc} V^{kl} \nabla_{e}V_{dj} + 24 
e^{2 \vp} R_{fgkl} V_{a}{}^{j} V_{b}{}^{k} V_{cd} 
\nabla_{e}V_{j}{}^{l} - 24 e^{2 \vp} R_{fgkl} V_{ab} 
V_{cd} V^{jk} \nabla_{e}V_{j}{}^{l}\nn\\&& + 12 e^{3 \vp} V_{a}{}^{j} 
V_{bc} V_{de} V_{j}{}^{k} V_{k}{}^{l} \nabla_{g}V_{fl} + 12 e^{3 
\vp} V_{a}{}^{j} V_{b}{}^{k} V_{cd} V_{ef} V_{j}{}^{l} 
\nabla_{g}V_{kl} - 24 e^{\vp} R_{dej}{}^{l} 
R_{fgkl} V_{a}{}^{j} \nabla^{k}V_{bc}\nn\\&& - 6 e^{2 \vp} 
V_{a}{}^{j} \nabla_{c}V_{bj} \nabla_{g}V_{fk} \nabla^{k}V_{de} - 
36 e^{2 \vp} R_{fgkl} V_{a}{}^{j} V_{bc} V_{j}{}^{k} 
\nabla^{l}V_{de},\labell{LW}
\eeqa
and $\cL^{\bar{H}}_{abcdef}$ contains 80 terms. They are
\beqa
\cL^{\bar{H}}_{abcdef}&\!\!\!=\!\!\!&2 e^{2 \vp} R_{cd}{}^{lm} R_{eflm} V_{a}{}^{j} 
V_{b}{}^{k} V_{jk} - 6 e^{3 \vp} R_{eflm} V_{a}{}^{j} 
V_{b}{}^{k} V_{c}{}^{l} V_{d}{}^{m} V_{jk} + 16 e^{\vp} 
R_{abj}{}^{l} R_{cdk}{}^{m} R_{eflm} 
V^{jk}\nn\\&& - 4 e^{\vp} R_{abjk} R_{cd}{}^{lm} 
R_{eflm} V^{jk} + e^{2 \vp} R_{cd}{}^{lm} 
R_{eflm} V_{ab} V_{jk} V^{jk} + 8 e^{3 \vp} 
R_{eflm} V_{a}{}^{j} V_{b}{}^{k} V_{cd} V_{j}{}^{l} 
V_{k}{}^{m} \nn\\&&+ 4 e^{3 \vp} R_{eflm} V_{ab} V_{cd} 
V_{j}{}^{l} V^{jk} V_{k}{}^{m} + \frac{3}{2} e^{4 \vp} 
V_{a}{}^{j} V_{b}{}^{k} V_{c}{}^{l} V_{d}{}^{m} V_{ef} V_{jk} V_{lm} 
\nn\\&&- 2 e^{4 \vp} V_{a}{}^{j} V_{b}{}^{k} V_{cd} V_{ef} V_{j}{}^{l} 
V_{k}{}^{m} V_{lm} -  \frac{1}{2} e^{4 \vp} V_{ab} V_{cd} 
V_{ef} V_{j}{}^{l} V^{jk} V_{k}{}^{m} V_{lm} \nn\\&&+ 2 e^{2 \vp} 
R_{cdlm} R_{efjk} V_{a}{}^{j} V_{b}{}^{k} V^{lm} - 
2 e^{3 \vp} R_{eflm} V_{a}{}^{j} V_{b}{}^{k} V_{cd} 
V_{jk} V^{lm}\nn\\&& - 12 e^{2 \vp} R_{cdjl} R_{efkm} 
V_{ab} V^{jk} V^{lm} + 2 e^{2 \vp} R_{cdjk} 
R_{eflm} V_{ab} V^{jk} V^{lm} \nn\\&&-  e^{3 \vp} 
R_{eflm} V_{ab} V_{cd} V_{jk} V^{jk} V^{lm} -  e^{3 \vp} 
R_{efjk} V_{a}{}^{j} V_{b}{}^{k} V_{cd} V_{lm} V^{lm}\labell{LH}\\&& + 
\frac{1}{2} e^{4 \vp} V_{a}{}^{j} V_{b}{}^{k} V_{cd} V_{ef} 
V_{jk} V_{lm} V^{lm} + \frac{1}{8} e^{4 \vp} V_{ab} V_{cd} 
V_{ef} V_{jk} V^{jk} V_{lm} V^{lm} \nn\\&&+ 16 e^{2 \vp} 
R_{efkl} V_{b}{}^{j} V_{c}{}^{k} \nabla_{a}\vp 
\nabla_{d}V_{j}{}^{l} - 32 e^{2 \vp} R_{efkl} V_{bc} 
V^{jk} \nabla_{a}\vp \nabla_{d}V_{j}{}^{l}\nn\\&& - 2 e^{3 \vp} 
V_{b}{}^{j} V_{cd} V_{kl} V^{kl} \nabla_{a}\vp \nabla_{f}V_{ej} 
- 12 e^{3 \vp} V_{b}{}^{j} V_{c}{}^{k} V_{d}{}^{l} V_{jk} 
\nabla_{a}\vp \nabla_{f}V_{el} \nn\\&&+ 16 e^{3 \vp} V_{b}{}^{j} 
V_{cd} V_{j}{}^{k} V_{k}{}^{l} \nabla_{a}\vp \nabla_{f}V_{el} + 
24 e^{3 \vp} V_{b}{}^{j} V_{c}{}^{k} V_{de} V_{j}{}^{l} 
\nabla_{a}\vp \nabla_{f}V_{kl} \nn\\&&+ 8 e^{3 \vp} V_{a}{}^{j} 
V_{b}{}^{k} V_{cd} \nabla_{e}V_{j}{}^{l} \nabla_{f}V_{kl} - 4 
e^{\vp} R_{cd}{}^{kl} R_{efkl} V_{bj} 
\nabla_{a}\vp \nabla^{j}\vp \nn\\&&+ 16 e^{\vp} 
R_{cdj}{}^{l} R_{efkl} V_{b}{}^{k} \nabla_{a}\vp \nabla^{j}\vp + 12 e^{2 \vp} R_{efkl} V_{bj} 
V_{c}{}^{k} V_{d}{}^{l} \nabla_{a}\vp \nabla^{j}\vp \nn\\&&- 8 
e^{2 \vp} R_{efkl} V_{b}{}^{k} V_{cd} V_{j}{}^{l} 
\nabla_{a}\vp \nabla^{j}\vp - 6 e^{3 \vp} V_{bj} 
V_{c}{}^{k} V_{d}{}^{l} V_{ef} V_{kl} \nabla_{a}\vp 
\nabla^{j}\vp \nn\\&&+ 4 e^{3 \vp} V_{b}{}^{k} V_{cd} V_{ef} 
V_{j}{}^{l} V_{kl} \nabla_{a}\vp \nabla^{j}\vp - 8 e^{2 
\vp} R_{efjl} V_{b}{}^{k} V_{cd} V_{k}{}^{l} \nabla_{a}
\vp \nabla^{j}\vp \nn\\&&+ 4 e^{2 \vp} R_{efkl} V_{bj} 
V_{cd} V^{kl} \nabla_{a}\vp \nabla^{j}\vp -  e^{3 \vp} 
V_{bj} V_{cd} V_{ef} V_{kl} V^{kl} \nabla_{a}\vp 
\nabla^{j}\vp \nn\\&&+ 8 e^{\vp} R_{cd}{}^{kl} 
R_{efkl} \nabla_{b}V_{aj} \nabla^{j}\vp - 8 e^{2 \vp} R_{efkl} V_{a}{}^{k} V_{b}{}^{l} \nabla_{d}V_{cj} 
\nabla^{j}\vp  \nn\\&&- 8 e^{2 \vp} R_{efkl} V_{ab} V^{kl} 
\nabla_{d}V_{cj} \nabla^{j}\vp+ 16 e^{2 \vp} 
R_{efjl} V_{a}{}^{k} V_{b}{}^{l} \nabla_{d}V_{ck} 
\nabla^{j}\vp  \nn\\&&+ 16 e^{2 \vp} R_{efjl} V_{ab} V^{kl} 
\nabla_{d}V_{ck} \nabla^{j}\vp + 4 e^{3 \vp} V_{a}{}^{k} 
V_{b}{}^{l} V_{cd} V_{kl} \nabla_{f}V_{ej} \nabla^{j}\vp  \nn\\&&+ 2 
e^{3 \vp} V_{ab} V_{cd} V_{kl} V^{kl} \nabla_{f}V_{ej} 
\nabla^{j}\vp + 16 e^{2 \vp} V_{b}{}^{k} \nabla_{a}\vp 
\nabla_{d}V_{cj} \nabla_{f}V_{ek} \nabla^{j}\vp  \nn\\&&+ 8 e^{3 \vp} V_{a}{}^{k} V_{b}{}^{l} V_{cd} V_{jk} \nabla_{f}V_{el} 
\nabla^{j}\vp  + 8 e^{3 \vp} V_{aj} V_{b}{}^{k} V_{cd} 
V_{k}{}^{l} \nabla_{f}V_{el} \nabla^{j}\vp \nn\\&&+ 8 e^{3 \vp} 
V_{ab} V_{cd} V_{j}{}^{k} V_{k}{}^{l} \nabla_{f}V_{el} 
\nabla^{j}\vp + 2 e^{\vp} R_{cd}{}^{kl} 
R_{efkl} V_{ab} \nabla_{j}\vp \nabla^{j}\vp \nn\\&&- 2 
e^{2 \vp} R_{efkl} V_{a}{}^{k} V_{b}{}^{l} V_{cd} 
\nabla_{j}\vp \nabla^{j}\vp + e^{3 \vp} V_{a}{}^{k} 
V_{b}{}^{l} V_{cd} V_{ef} V_{kl} \nabla_{j}\vp \nabla^{j}\vp \nn\\&&- 2 e^{2 \vp} R_{efkl} V_{ab} V_{cd} V^{kl} 
\nabla_{j}\vp \nabla^{j}\vp + \frac{1}{2} e^{3 \vp} 
V_{ab} V_{cd} V_{ef} V_{kl} V^{kl} \nabla_{j}\vp 
\nabla^{j}\vp \nn\\&&+ 4 e^{2 \vp} V_{b}{}^{k} V_{cd} 
\nabla_{a}\vp \nabla_{f}V_{ek} \nabla_{j}\vp 
\nabla^{j}\vp - 16 e^{\vp} R_{cdj}{}^{l} 
R_{efkl} \nabla^{j}\vp \nabla^{k}V_{ab} \nn\\&&- 8 e^{2 \vp} R_{efjl} V_{a}{}^{j} \nabla_{d}V_{k}{}^{l} 
\nabla^{k}V_{bc} - 16 e^{2 \vp} R_{efkl} V_{a}{}^{j} 
\nabla_{b}V_{j}{}^{l} \nabla^{k}V_{cd} \nn\\&&+ 4 e^{2 \vp} V_{bj} 
\nabla_{a}\vp \nabla_{f}V_{ek} \nabla^{j}\vp 
\nabla^{k}V_{cd} - 8 e^{2 \vp} \nabla_{b}V_{aj} 
\nabla_{f}V_{ek} \nabla^{j}\vp \nabla^{k}V_{cd} \nn\\&&- 2 e^{2 \vp} V_{ab} \nabla_{f}V_{ek} \nabla_{j}\vp \nabla^{j}\vp 
\nabla^{k}V_{cd} - 16 e^{2 \vp} R_{efjl} V_{ab} 
\nabla_{d}V_{k}{}^{l} \nabla^{k}V_{c}{}^{j}\nn\\&& - 8 e^{\vp} 
R_{cdj}{}^{l} R_{efkl} V_{ab} \nabla^{j}\vp 
\nabla^{k}\vp + 8 e^{2 \vp} R_{efkl} V_{aj} 
V_{b}{}^{l} V_{cd} \nabla^{j}\vp \nabla^{k}\vp \nn\\&&+ 8 e^{2 
\vp} R_{efkl} V_{ab} V_{cd} V_{j}{}^{l} \nabla^{j}\vp \nabla^{k}\vp - 4 e^{3 \vp} V_{aj} V_{b}{}^{l} V_{cd} 
V_{ef} V_{kl} \nabla^{j}\vp \nabla^{k}\vp\nn\\&& - 2 e^{3 \vp} 
V_{ab} V_{cd} V_{ef} V_{j}{}^{l} V_{kl} \nabla^{j}\vp 
\nabla^{k}\vp + 8 e^{2 \vp} V_{bj} V_{cd} \nabla_{a}\vp 
\nabla_{f}V_{ek} \nabla^{j}\vp \nabla^{k}\vp \nn\\&&- 8 e^{2 
\vp} V_{ab} \nabla_{d}V_{cj} \nabla_{f}V_{ek} \nabla^{j}\vp \nabla^{k}\vp + 2 e^{2 \vp} V_{bk} V_{cd} V_{ef} 
\nabla_{a}\vp \nabla_{j}\vp \nabla^{j}\vp 
\nabla^{k}\vp\nn\\&& - 4 e^{2 \vp} V_{ab} V_{cd} \nabla_{f}V_{ek} 
\nabla_{j}\vp \nabla^{j}\vp \nabla^{k}\vp -  
\frac{1}{2} e^{2 \vp} V_{ab} V_{cd} V_{ef} \nabla_{j}\vp 
\nabla^{j}\vp \nabla_{k}\vp \nabla^{k}\vp \nn\\&&- 8 e^{3 
\vp} V_{a}{}^{j} V_{b}{}^{k} V_{j}{}^{l} \nabla_{d}V_{ck} 
\nabla_{l}V_{ef} - 2 e^{2 \vp} R_{efjk} V^{jk} 
\nabla_{d}V_{cl} \nabla^{l}V_{ab}\nn\\&& - 32 e^{2 \vp} 
R_{efkl} V_{b}{}^{j} V_{j}{}^{k} \nabla_{a}\vp 
\nabla^{l}V_{cd} + 8 e^{2 \vp} R_{efkl} V^{jk} 
\nabla_{b}V_{aj} \nabla^{l}V_{cd}\nn\\&& - 8 e^{2 \vp} 
R_{efkl} V_{aj} V_{b}{}^{k} \nabla^{j}\vp 
\nabla^{l}V_{cd} - 8 e^{2 \vp} R_{efkl} V_{ab} 
V_{j}{}^{k} \nabla^{j}\vp \nabla^{l}V_{cd}\nn\\&& + e^{3 \vp} 
V_{a}{}^{j} V_{b}{}^{k} V_{jk} \nabla_{l}V_{ef} \nabla^{l}V_{cd} + 
\frac{1}{2} e^{3 \vp} V_{ab} V_{jk} V^{jk} \nabla_{l}V_{ef} 
\nabla^{l}V_{cd} - 16 e^{2 \vp} R_{efkl} V_{ab} 
\nabla^{k}V_{c}{}^{j} \nabla^{l}V_{dj} \nn\\&&+ 16 e^{2 \vp} 
R_{efjl} V_{ab} \nabla^{k}V_{c}{}^{j} \nabla^{l}V_{dk} - 8 
e^{2 \vp} R_{efkl} V_{a}{}^{j} \nabla_{c}V_{bj} 
\nabla^{l}V_{d}{}^{k} - 8 e^{3 \vp} V_{a}{}^{j} V_{bc} 
V_{j}{}^{k} \nabla_{d}V_{kl} \nabla^{l}V_{ef}\nn.
\eeqa
In above equations $R_{abcd}$ is the Riemann curvature made of the base space metric $\bg_{ab}$, $V_{ab}=\prt_a g_b-\prt_b g_a$ and $W_{ab}=\prt_a b_b-\prt_b b_a$ are the field strengths of the verctors, and base space torsion $\bH_{abc}$ is \cite{Kaloper:1997ux}
\beqa
\bH_{abc}=3\prt_{[a}\tilde{b}_{bc]}-3W_{[ab}\,g_{c]}
\eeqa
where $\tilde{b}_{ab}=\bb_{ab} + b_{[a} g_{b]}$. It is important to note that for a zero RR field, the leading two-derivative type IIA action contains no Chern-Simons term. Consequently, there is no eight-derivative term generated by a six-derivative field redefinition of the leading-order action. As a result, the field redefinition cannot alter the form of the couplings in \reef{reduce1}.

The action in equation \reef{reduce1}, which captures the one-loop effective action of type IIA theory in the large-radius limit, can also be derived from the relevant torus-level S-matrix elements. In type IIA string theory compactified on a circle, the continuous momentum along the circle is replaced by discrete KK and winding momenta \cite{Green:1982sw}. Consequently, the corresponding spacetime couplings generally receive contributions from both types of modes. The couplings in \reef{reduce1}, however, arise exclusively from the pure KK sector. In the next section, we will use T-duality to derive from these type IIA couplings the corresponding pure winding contributions for type IIB theory.
        
\subsection{9D Chern-Simons term in type IIB }

Under T-duality, the pure KK modes in type IIA S-matrix elements become the pure winding modes in type IIB. Consequently, the couplings in \reef{reduce1}—which correspond to these pure KK modes and are valid in the large-radius limit—transform into couplings for the pure winding modes, valid in the small-radius limit. 

Under T-duality, the base space fields transform as \cite{Buscher:1987sk,Rocek:1991ps,Kaloper:1997ux}:
\beqa
\bg_{ab}'=\bg_{ab}\,,\bH_{abc}'=\bH_{abc}\,,\bphi'=\bphi\,,g_a'=b_a\,,b'_a=g_a\,,\vp'=-\vp\,.\labell{T-dual}
\eeqa
Therefore, under T-duality, the nine-dimensional type IIA couplings given in \reef{reduce1} yield the following type IIB couplings:
\beqa
S^{CS}_{\rm IIB}&\!\!\!=\!\!\!&-\frac{2}{\kappa^2}\frac{\pi^3\alpha'^3}{2^{10}.3^2}\int d^{9}x\sqrt{-\bg} \,\epsilon_{9}^{abcdefghi}\Big[g_a\cL_{ bcdefghi}\!+\!V_{hi}\cL'^W_{ abcdefg }\!+\!\bar{H}_{ghi}\cL'^{\bar{H}}_{abcdef}\Big]\!,\labell{reduce2}
\eeqa
where $\cL'^W$ and $\cL'^{\bar{H}}$ are identical to the expressions in \reef{LW} and \reef{LH}, but with $V$ replaced by $W$ and the sign of $\vp$ reversed. 
The resulting action provides the correct couplings for type IIB theory in the small-radius limit. Note that the action contains only even numbers of \(W\) or \(\bH\). As a feature of type IIB theory, these couplings must be consistent with S-duality; we will examine this consistency in the next section.

\subsubsection{Consistency with S-duality}

Unlike the T-duality transformations in \reef{T-dual}, which are given in the string frame, the S-duality transformations of type IIB theory are properly defined in the Einstein frame. In this frame, the metric and the RR four-form are invariant, while the B-field and the RR two-form transform as a doublet \cite{Tseytlin:1996it,Green:1996qg}. Since the duality parameters are constant, their field strengths also transform as a doublet, \ie 
\beqa
\cH\equiv\pmatrix{H \cr 
F}\rightarrow (\Lambda^{-1})^T \pmatrix{H \cr 
F}\,\,\,;\,\,\,\Lambda=\pmatrix{p&q \cr 
r&s}\in SL(2,R)\,.\labell{2}
\eeqa
The dilaton and the RR scalar transform nonlinearly as $\tau \rightarrow \frac{p\tau + q}{r\tau + s}$, where the complex scalar field is defined as $\tau = C + i e^{-\Phi}$. The matrix $\mathcal{M}$, defined in terms of the dilaton and the RR scalar, \ie 
 \beqa
 {\cal M}=e^{\Phi}\pmatrix{|\tau|^2&C \cr 
C&1}\labell{M}
\eeqa
then  transforms  as \cite{Gibbons:1995ap}
\beqa
{\cal M}\rightarrow \Lambda {\cal M}\Lambda ^T
\eeqa
The derivatives of this matrix transform in the same way.  The matrix $\cN$, which is defined as 
 \beqa
 {\cal N}=\pmatrix{0&1 \cr 
-1&0}\labell{N}
\eeqa
satisfies $
\Lambda {\cal N}\Lambda ^T$. Using these matrices and the transformation in \reef{2}, one can construct various $SL(2,\MR)$-invariant objects. For example, $\cH^T_{\mu\nu\alpha}{\cal N}\cH_{\beta\gamma\lambda}=-F_{\mu\nu\alpha}H_{\beta\gamma\lambda}+H_{\mu\nu\alpha}F_{\beta\gamma\lambda}$ is invariant under the $SL(2,\MR)$ transformation.

 An \(SL(2,\MR)\)-invariant object, in general, has more than one component. For example, \(\cH^T\cM\cH\) has the following components:
 \beqa
 \cH^T\cM\cH = e^{-\Phi}(1 + e^{2\Phi}C^2)HH + e^{\Phi}FF - e^{\Phi}C(HF + FH).\labell{SHH}
 \eeqa
 When the RR fields  are zero, the following terms are invariant under S-duality:
 \beqa
 e^{-\Phi}H_{\mu\nu\alpha}H_{\beta\gamma\lambda} \rightarrow e^{-\Phi}H_{\mu\nu\alpha}H_{\beta\gamma\lambda}, && e^{-\Phi}W_{ab}W_{cd} \rightarrow e^{-\Phi}W_{ab}W_{cd}.\labell{S-dual}
 \eeqa
 Higher derivatives of these field strengths are also invariant.

Hence, to study the S-duality of the couplings, it is appropriate to transform the string frame metric to the Einstein frame metric using \(G_{\mu\nu} = e^{\Phi/2}G^E_{\mu\nu}\). Applying this transformation to the circularly reduced metric in \reef{reduc}, one finds the following relations between the string frame and Einstein frame base space fields:
\beqa
\bg_{ab} = e^{\Phi/2}\bg_{ab}^E\,,\quad g_a = g_a^E\,,\quad \vp = \frac{1}{2}\Phi + \vp^E.\labell{Evp}
\eeqa
Under S-duality, the ten-dimensional Einstein frame metric is invariant; hence, 
\beqa
\bg_{ab}^E\rightarrow \bg_{ab}^E\,,\quad g_a\rightarrow g_a\,,\quad \vp^E\rightarrow \vp^E\,.\labell{ggp}
\eeqa
 The Levi-Civita tensor \(\epsilon_{9}\) is related to the Levi-Civita symbol \(\epsilon'_{9}\) by \(\sqrt{-\bg}\,\epsilon_{9} = \epsilon'_{9}\). Therefore, the overall factor in \reef{reduce2} is invariant under S-duality.

We now continue the discussion for a constant dilaton in type IIB theory. It is then obvious that the first term in \reef{reduce2} is S-duality invariant, because for a constant dilaton one finds
\beqa
R_{abcd} = e^{\Phi/2}R^E_{abcd},
\eeqa
and the Lagrangian density \(\mathcal{L}_{bcdefghi}\) contains four Riemann tensors and four inverse metrics. We also confirmed that the dilaton derivative does not appear when transforming the first term in \reef{reduce2} to the Einstein frame.

The second term in \reef{reduce2} is also invariant under S-duality for zero RR field and a constant dilaton. Here, \( V_{hi} \) is invariant, and \( \mathcal{L}'^W_{abcdefg} \) is invariant as well. To see this, consider the first term in \( \mathcal{L}'^W_{abcdefg} \):
\beqa
-24 e^{-2 \vp} R_{fgkl} W_{b}{}^{j} W_{c}{}^{k} W_{de} W_{j}{}^{l} \nabla_{a}\vp.
\eeqa
This term contains one Riemann tensor and three inverse metrics. Hence, in the Einstein frame, it becomes
\beqa
-24 e^{-2 \vp^E} e^{-2\Phi} R^E_{fgkl} W_{b}{}^{j} W_{c}{}^{k} W_{de} W_{j}{}^{l} \nabla_{a}\vp^E,
\eeqa
which is invariant under S-duality according to \reef{S-dual}. Similarly, all other terms in \( \mathcal{L}'^W_{abcdefg}\) are S-duality invariant.

To study the S-duality of the third term in \reef{reduce2}, we first note that the following structure is invariant under S-duality for zero RR field:
\beqa
e^{-\Phi}\bH_{abc}W_{de} \rightarrow e^{-\Phi}\bH_{abc}W_{de}.
\eeqa
To see the invariance of the third term in \reef{reduce2}, consider, for example, the first term in \(\bar{H}_{ghi}\cL'^{\bar{H}}_{abcdef}\), which is given by
\beqa
2 e^{-2 \vp} R_{cd}{}^{lm} R_{eflm} \bar{H}_{ghi}W_{a}{}^{j} W_{b}{}^{k} W_{jk}.
\eeqa
This term contains two Riemann tensors and four inverse metrics. Hence, it transforms into the following coupling in the Einstein frame:
\beqa
2 e^{-2 \vp^E} e^{-2\Phi} R^E_{cd}{}^{lm} R^E_{eflm} \bar{H}_{ghi}W_{a}{}^{j} W_{b}{}^{k} W_{jk},
\eeqa
which is obviously invariant under S-duality. Similarly, all other terms in \(\bar{H}_{ghi}\cL'^{\bar{H}}_{abcdef}\) are invariant.

  Therefore, the type IIB couplings given in \reef{reduce2}, which are valid in the small-radius limit, are invariant under the continuous \( SL(2,\mathbb{R}) \) S-duality group—much like the two-derivative effective action of type IIB (see, e.g., \cite{Becker:2007zj}). That is,
 \beqa
 S_{\rm IIB}^{CS} &\rightarrow S_{\rm IIB}^{CS}.
 \eeqa
 In contrast, a non-zero ten-dimensional Chern–Simons term involving an even number of \(B\)-fields has been proposed in \cite{Liu:2013dna} and shown to be S-duality invariant after including tree-level couplings with RR fields as well as non-perturbative contributions \cite{Liu:2019ses}. The KK reduction of these terms yields S-duality invariant couplings in the large-radius limit. Consequently, while the small-radius couplings are invariant under the continuous \( SL(2,\mathbb{R}) \) group, the large-radius couplings become invariant under the discrete \( SL(2,\mathbb{Z}) \) group only after the inclusion of tree-level and non-perturbative corrections.
Since the tree-level and non-perturbative sectors are incorporated independently at large radius, and there are no two-loop or higher corrections to couplings at this derivative order \cite{Green:1998by,Berkovits:1997pj,Basu:2011he,Bossard:2014lra,DHoker:2005vch}, S-duality forces any additional one-loop terms to be separately invariant. This condition, together with the transformation of the radion field \(\varphi\) from the string frame to the Einstein frame (see \reef{Evp}), leads to the conclusion that only the small-radius coupling satisfies the required invariance. The same conclusion can also be reached by studying the duality in five dimensions, which we consider in the next section.

\subsection{Reduction  on K3 and duality in 5D}

Type IIB string theory compactified on $S^{1} \times K3 $ is known to be dual to heterotic string theory on $T^5$ (see, e.g., \cite{Becker:2007zj}). Each theory contains 106 scalar fields; however, we are interested in only one specific scalar from each—the dilaton in the heterotic theory and the radius of the circle in the type IIB theory. Both theories also contain 27 vector fields.
The type IIB string theory compactified on K3 yields 21 tensor multiplets and one gravity multiplet. Each tensor multiplet contains one two-form field, while the gravity multiplet includes five two-form fields and one metric. Upon further compactification on an additional circle, these fields give rise to 27 vector fields.
In the heterotic theory, the 27 vector fields arise from: five gauge fields derived from the metric upon compactification on $T^5$,
sixteen gauge fields from the Cartan subalgebra of $SO(32) $ or $E_8 \times E_8$,
five gauge fields from the reduction of the B-field on $T^5$,
and one vector field obtained via Hodge dualization of the B-field in five-dimensional spacetime.
 Although all 27 vectors are present in each theory, we focus specifically on two of them. In type IIB, we consider \(W\) and \(V\), which appear in the action \reef{reduce2}. In the heterotic theory, we consider one vector resulting from the metric along one circle of \(T^5\), and the other from the \(B\)-field along the same circle.
The field content of both theories also includes a five-dimensional metric and a Kalb-Ramond field.

The K3 reduction of the eight-derivative couplings in \reef{reduce2} generates both the four-derivative couplings in which we are interested and higher-derivative couplings in which we are not. To isolate the four-derivative couplings, we consider an ansatz where the 9-dimensional metric takes the block-diagonal form:
\beqa
ds^2 = G^5_{ab}(x)dx^a dx^b + g^4_{\mu\nu}(y)dy^\mu dy^\nu,\label{K3red}
\eeqa
with \(y^\mu\) denoting the K3 coordinates. In this section, we use the indices \(a,b,\cdots\) for the 5-dimensional space and the indices \(\mu,\nu,\cdots\) for the compact 4-dimensional space. For the block-diagonal metric, the 9-dimensional Levi-Civita symbol can be written as the product of the 5-dimensional and 4-dimensional Levi-Civita symbols. In terms of the Levi-Civita tensor, this is expressed as:
\beqa
\sqrt{-\bg}\,\epsilon_{9} = \sqrt{-G^5}\,\epsilon_{5}\ \sqrt{g^4}\,\epsilon_{4}.
\eeqa
The non-flatness of the K3 surface introduces non-vanishing curvature contributions. In particular, the integral of the first Pontryagin class over K3 is (see, e.g., \cite{Liu:2019ses}):
\beqa
\frac{1}{32\pi^2}\int_{K3} d^4y\sqrt{g^4}\,\epsilon_4^{\alpha\beta\mu\nu}R_{\alpha\beta \gamma\delta}R_{\mu\nu}{}^{\gamma\delta} = 48.\labell{K3int}
\eeqa
This topological constraint plays a key role in producing the four-derivative couplings when applied to the eight-derivative couplings in \reef{reduce2}.

Using the above constraint, the K3 reduction of the nine-dimensional, one-loop gravity couplings in \reef{reduce2} produces the following four-derivative term in five dimensions:
\beqa
S_{5D}&\!\!\!\!=\!\!\!\!&-\frac{2}{\kappa^2}\frac{\pi^5\alpha'^3}{6}\int d^{5}x\sqrt{-G^5}\,\epsilon_5^{abcde}\Big[-12g_a R_{bc}{}^{fg} R_{defg} +2
\bH_{cde} W_{a}{}^{f} W_{b}{}^{g} W_{fg}e^{-2 \vp}\nn\\&& - 4
\bH_{cde} R_{abfg} W^{fg}e^{-\vp} + \bH_{cde}
W_{ab} W_{fg} W^{fg}e^{-2 \vp} + 12 V_{de} W_{a}{}^{f}
\nabla_{c}W_{bf}e^{-\vp}\nn\\&& - 4 \bH_{cde} W_{bf}
\nabla_{a}\vp \nabla^{f}\vp e^{-\vp} - 8
\bH_{cde} \nabla_{b}W_{af} \nabla^{f}\vp e^{-\vp} +
2 \bH_{cde} W_{ab} \nabla_{f}\vp \nabla^{f}\vp e^{-\vp}\Big].\labell{SR2}
\eeqa
To convert the first term into the one that appears in the heterotic theory, we use the following identity:
\beqa
\frac{1}{4}\epsilon_5^{abcde}R_{fgab}R^{fg}{}_{cd}&=&\epsilon_5^{abcde}\nabla_{d}\Omega_{abc},
\eeqa
where \(\Omega_{abc}\) is the Lorentz Chern-Simons three-form in five dimensions.
Then the first term in \reef{SR2} can be written in terms of \(\epsilon_5^{abcde}V_{ab}\Omega_{cde}\) up to a total derivative.     Note that since the form of the couplings in the 9-dimensional action \reef{reduce1} and its T-dual action \reef{reduce2} does not change under higher-derivative field redefinitions, the corresponding 5-dimensional couplings in \reef{SR2} also remain unchanged.

We now map the five-dimensional one-loop type IIB theory \reef{SR2} to the couplings in the five-dimensional heterotic theory using the following map:
\beqa
&&\bH^{abc}\rightarrow\frac{e^{-\Phi}}{2!}\epsilon_5^{abcde}W_{de}\,,\,V^{ab}\rightarrow-\frac{e^{-3\Phi}}{3!}\epsilon_5^{abcde}\bH_{cde}\,,\,W_{ab}\rightarrow V_{ab}\,,\nn\\&&G_{ab}\rightarrow e^{-2\Phi}G_{ab}\,,\,\vp\rightarrow 2\Phi\,,\,\Omega_{abc}\,\rightarrow\,\bar{\Omega}_{abc},\labell{map}
\eeqa
where the fields on the right-hand sides are those of the dual theory. Note that the radius field \(\vp\) in type IIB maps to the dilaton in the heterotic theory. Under the above transformation, the five-dimensional action \reef{SR2} transforms into the following dual action:
\beqa
S^{\rm dual}_{5D}&=&-\frac{2}{\kappa^2}\frac{\pi^5\alpha'^3}{6}\int d^{5}x\sqrt{-G^5}\,e^{-2\Phi}\Big[ 48 \bH^{abc} \bar{\Omega}_{abc} -12 V_{a}{}^{c} V^{ab} V_{b}{}^{d} W_{cd} - 6 V_{ab} V^{ab} V^{cd}
W_{cd}  \nn\\&&+ 24 R_{abcd} V^{ab} W^{cd} - 48 \bH_{bcd} V_{a}{}^{b} V^{cd} \nabla^{a}\Phi - 96
W^{bc} \nabla^{a}\Phi \nabla_{c}V_{ab} + 96 V^{ab} W_{a}{}^{c}
\nabla_{c}\nabla_{b}\Phi \nn\\&&+ 24 \bH_{bcd} V^{ab}
\nabla^{d}V_{a}{}^{c}\Big].\labell{SR21}
\eeqa
Note that the overall dilaton factor \(e^{-2\Phi}\) indicates the dual action is at the sphere level. Note also that the above action is linear in the field strengths \(\bar{H}\) and \(W\), so it should correspond to couplings in the heterotic theory that are linear in the NS-NS antisymmetric tensor field strength.

Using the fact that under the five-dimensional duality between type IIB and heterotic theory, the NS5-brane of heterotic theory wrapped on \(T^5\) with volume \(V\) transforms into the fundamental string of type IIB theory wrapped on the circle, the equality of their masses yields the following relation when the volume of the circle is \(2\pi\):
\beqa
\frac{2\pi V}{(2\pi\sqrt{\alpha'})^6 g_s^2} = \frac{2\pi}{2\pi\alpha'}.\labell{NS5F}
\eeqa
Using this relation, one finds that \reef{SR21} produces exactly the reduction of the Chern-Simons term \(H^{\mu\nu\alpha}\Omega_{\mu\nu\alpha}\) in the heterotic theory on \(T^5\). This holds when the metric and \(B\)-field have a non-zero vector component along one circle of fixed radius, and after applying an appropriate two-derivative field redefinition.  This calculation closely follows that of \cite{Garousi:2025dil}, which demonstrated that the K3 reduction of analogous ten-dimensional type IIA couplings yields the reduction of the Chern-Simons term \(H^{\mu\nu\alpha}\Omega_{\mu\nu\alpha}\) in the heterotic theory on \(T^4\). The duality between IIA theory on K3 and heterotic theory on \(T^4\) has also been studied at the four-derivative order in \cite{Liu:2013dna}. Additionally, the supersymmetrization of the four-derivative couplings and their consistency between type II and heterotic string theories are examined in \cite{Chang:2022urm}.

It is well-known that the four-derivative NS-NS couplings in the heterotic theory do not receive higher-genus corrections \cite{Ellis:1987dc,Ellis:1989fi}. Then the map \reef{map}, in which the radius maps to the dilaton, indicates that the radius dependence in \reef{SR2} is exact. Hence, the radius dependence in the 9-dimensional action \reef{reduce2} is also exact, which is consistent with the conclusion reached from S-duality in subsection 2.3.1.

\section{Pure gravity couplings  at order $\alpha'^3$}

The KK reduction of the pure gravity couplings in M-theory yields the corresponding pure gravity couplings in type IIA theory, along with other couplings involving RR fields \cite{Aggarwal:2025lxf,Garousi:2025wfk} that are not of interest here. The pure gravity couplings in type IIA theory are
\beqa
\bS_{\rm IIA}&=&-\frac{2}{\kappa^2}\frac{\pi^2\alpha'^3}{2^{3}.3}\int d^{10}x\sqrt{-G}\Big[\frac{1}{4} R_{\alpha \beta }{}^{\epsilon \varepsilon 
} R^{\alpha \beta \gamma \delta } R_{\gamma 
\epsilon }{}^{\mu \nu } R_{\delta \varepsilon \mu \nu 
} -  R_{\alpha \beta }{}^{\epsilon \varepsilon } 
R^{\alpha \beta \gamma \delta } R_{\gamma 
}{}^{\mu }{}_{\epsilon }{}^{\nu } R_{\delta \mu 
\varepsilon \nu }\labell{ttee1}\\&&\qquad\qquad\qquad\qquad\qquad + \frac{1}{16} R_{\alpha \beta }{}^{
\epsilon \varepsilon } R^{\alpha \beta \gamma \delta } 
R_{\gamma \delta }{}^{\mu \nu } R_{\epsilon 
\varepsilon \mu \nu } + \frac{1}{2} R_{\alpha 
}{}^{\epsilon }{}_{\gamma }{}^{\varepsilon } R^{\alpha 
\beta \gamma \delta } R_{\beta }{}^{\mu }{}_{\delta 
}{}^{\nu } R_{\epsilon \mu \varepsilon \nu }\nn\\&&\qquad\qquad\qquad\qquad\qquad -  
R_{\alpha \beta \gamma }{}^{\epsilon } 
R^{\alpha \beta \gamma \delta } R_{\delta 
}{}^{\varepsilon \mu \nu } R_{\epsilon \mu \varepsilon 
\nu } + \frac{1}{32} R_{\alpha \beta \gamma \delta } 
R^{\alpha \beta \gamma \delta } R_{\epsilon 
\varepsilon \mu \nu } R^{\epsilon \varepsilon \mu \nu 
}\Big].\nn
\eeqa
Up to field redefinitions, it is the familiar coupling $(t_8t_8-\frac{1}{8}\epsilon_{10}\epsilon_{10})R^4$ \cite{Sakai:1986bi,Russo:1997mk,Vafa:1995fj,Antoniadis:1997eg}. There is no dilaton in the above action; therefore, it is the one-loop effective action of type IIA string theory.

In the next subsection, we find the KK reduction of the above coupling. Since the above action is the one-loop effective action of type IIA string theory, the KK reduction does not produce the complete couplings in nine dimensions. There should be winding contributions as well.  These can be found by applying T-duality to the KK reduction of the one-loop effective action of type IIB theory, which is given by
\beqa
\bS_{\rm IIB}&=&-\frac{2}{\kappa^2}\frac{\pi^2\alpha'^3}{2^{3}.3}\int d^{10}x\sqrt{-G}\Big[  R_{\alpha }{}^{\epsilon 
}{}_{\gamma }{}^{\varepsilon } R^{\alpha \beta \gamma 
\delta } R_{\beta }{}^{\mu }{}_{\epsilon }{}^{\nu } 
R_{\delta \mu \varepsilon \nu } - \frac{1}{4} R_{\alpha \beta }{}^{\epsilon 
\varepsilon } R^{\alpha \beta \gamma \delta } 
R_{\gamma \epsilon }{}^{\mu \nu } R_{\delta 
\varepsilon \mu \nu }\Big].\labell{IIB}
\eeqa
Up to field redefinition, it is the coupling $(t_8t_8+\frac{1}{8}\epsilon_{10}\epsilon_{10})R^4$ \cite{Sakai:1986bi,Russo:1997mk,Vafa:1995fj,Antoniadis:1997eg}.  Applying KK reduction and then T-duality to the above couplings leads to the type IIA winding-mode couplings in the small-radius limit. In M-theory, these are interpreted as coming from an M2-brane wrapped on \( S^{1} \times S^{1} \). We will not explore this correspondence further in the present work.

The same couplings as above appear at tree-level in type IIB theory  \cite{Gross:1986iv,Gross:1986mw,Grisaru:1986kw,Grisaru:1986vi}. The combination of tree-level, one-loop, and non-perturbative effects causes them to be invariant under the S-duality of type IIB theory in ten dimensions \cite{Green:1997tv}. Consequently, their circular KK reduction should also satisfy S-duality in nine dimensions \cite{Green:1997di}. The KK reduction is valid only in the large-radius limit.  S-duality then requires that if there are couplings in the nine-dimensional type IIB theory at the small-radius limit, they must satisfy S-duality by themselves.

To find the nine-dimensional couplings for type IIB theory in the small-radius limit, one should first perform the KK reduction of the type IIA couplings in \reef{ttee1}. Since the result of this reduction is valid only in the large-radius limit, one must then apply T-duality to obtain the desired couplings in type IIB theory. In the following section, we calculate this KK reduction of the type IIA couplings given in \reef{ttee1}.

\subsection{Large-radius limit of 9D couplings in type IIA }

The dimensional reduction scheme in \reef{reduc} yields the following nine-dimensional type IIA couplings:
\beqa
S_{IIA}&=&-\frac{2}{\kappa^2}\frac{\pi^3\alpha'^3}{2^{2}.3}\int d^{9}x\sqrt{-\bg}\,e^{\vp/2}\Big[\frac{1}{4} R_{a b }{}^{d e 
} R^{a bc f } R_{c d }{}^{g h } R_{f e g h 
} -  R_{a b }{}^{d e } 
R^{a bc f } R_{c 
}{}^{g }{}_{d }{}^{h } R_{f g 
e h }\labell{ttee12}\\&&\qquad\qquad\qquad\qquad\qquad + \frac{1}{16} R_{a b }{}^{
d e } R^{a bc f } 
R_{c f }{}^{g h } R_{d 
e g h } + \frac{1}{2} R_{a 
}{}^{d }{}_{c }{}^{e } R^{a 
bc f } R_{b }{}^{g }{}_{f 
}{}^{h } R_{d g e h }\nn\\&&\qquad\qquad\qquad\qquad\qquad -  
R_{a bc }{}^{d } 
R^{a bc f } R_{f 
}{}^{e g h } R_{d g e 
h } + \frac{1}{32} R_{a bc f } 
R^{a bc f } R_{d 
e g h } R^{d e g h 
}+\cdots\Big].\nn
\eeqa
 The ellipsis in the equation represents 708 couplings involving the Riemann, Ricci, and Ricci scalar curvatures, along with first and second derivatives of the radius parameter \( \varphi \) and the \( U(1) \) gauge field \( g_a \).

The couplings in \reef{ttee12} represent one-loop effective interactions in a specific scheme. While these can be transformed into alternative schemes via field redefinitions and integration by parts, constructing a representation with the minimal number of couplings presents a non-trivial challenge. Although we cannot determine the absolute minimum number of independent couplings, in this section we express them using a minimal basis of 298 eight-derivative metric-radion-vector interactions. As detailed in the Appendix, this minimal basis admits multiple equivalent representations. Here we demonstrate that the couplings in \reef{ttee12} can be expanded in this basis with 288 non-zero coefficients and 10 vanishing coefficients. The specific values of these coefficients are scheme-dependent, contingent upon the chosen representation of the minimal basis.

 We  present the final result for the one-loop effective action of type IIA theory at eight-derivative order, restricted to the metric-radion-vector sector:
\beqa
S_{IIA}&=&-\frac{2}{\kappa^2}\frac{\pi^3\alpha'^3}{2^{2}.3}\int d^{9}x\sqrt{-\bg}\,\cL(\vp,V)\,,\labell{ttee3}
\eeqa 
where the Lagrangian \(\mathcal{L}(\varphi,V)\) is given in \reef{T55}, with the coupling constants set to the values in \reef{sol}. Since this action is derived from the KK reduction of the ten-dimensional couplings in \reef{ttee1}, it is valid only in the large-radius limit. The corresponding couplings in type IIA in the small-radius limit can be found via T-duality from the KK reduction of the type IIB couplings in \reef{IIB}, which is not the focus of this work. In the next section, we instead derive the small-radius limit of the type IIB couplings  at order \(\alpha'^3\).

\subsection{Small-radius limit of 9D couplings in type IIB }

Under T-duality, the effective action corresponding to the pure KK modes in type IIA theory becomes the effective action of type IIB that corresponds to pure winding modes.

Under T-duality \reef{T-dual}, the nine-dimensional type IIA couplings given in \reef{ttee3} yield the following type IIB couplings in nine dimensions:
\beqa
S_{IIB}&=&-\frac{2}{\kappa^2}\frac{\pi^3\alpha'^3}{2^{2}.3}\int d^{9}x\sqrt{-\bg}\,\cL(-\vp,W)\,,\labell{ttee31}
\eeqa 
where \(\mathcal{L}(-\varphi,W)\) is identical to \(\mathcal{L}(\varphi,V)\) but with \(V\) replaced by \(W\) and the sign of \(\varphi\) reversed. The resulting action represents the correct couplings of type IIB theory in the small-radius limit.

As these couplings belong to the type IIB framework, they are expected to be consistent with S-duality. In the large-radius limit of type IIB, achieving S-duality invariance—which is governed by the discrete \( SL(2,\mathbb{Z}) \) group—requires the inclusion of tree-level and non-perturbative effects \cite{Green:1997tv}. Conversely, since the tree-level effective action contains no winding modes, no tree-level action is expected to exist in the small-radius limit.\footnote{It is important to clarify that the KK reduction of the ten-dimensional tree-level action \( \sqrt{-G} e^{-2\Phi} (R - \frac{1}{12}H^2) \), assuming a constant dilaton, gives
 \[
 \sqrt{-\bg} e^{-2\bphi} \Bigl( R - \frac{1}{4} \nabla_a \varphi \nabla^a \varphi - \frac{e^{\varphi}}{4} V^2 - \frac{e^{-\varphi}}{4} W^2 - \frac{1}{12} \bH^2 \Bigr).
 \]
 Although this expression includes both \( e^{\varphi} \) and \( e^{-\varphi} \), it corresponds entirely to the large-radius regime. This is because it originates from the standard dimensional reduction and does not include contributions from winding modes, which characterize the small-radius limit.
} Therefore, the one-loop couplings at the small-radius limit must be S-duality invariant by themselves, corresponding to the continuous \( SL(2,\mathbb{R}) \) symmetry. In the next section, we study the S-duality of the couplings in \reef{ttee31}.

\subsubsection{Consistency with S-duality}

    To analyze the S-duality properties of the couplings, we must first express them in the Einstein frame, as in subsection 2.3.1. We begin by considering the pure gravity couplings given in \reef{ttee31}:
\beqa
\sqrt{-\bg}\,e^{-\vp/2}RRRR
\eeqa
The transformation of \(\sqrt{-\bg}\) to the Einstein frame is \(e^{9\Phi/4}\sqrt{-\bg^E}\). Similarly, \(e^{-\vp/2}\) transforms to \(e^{-\Phi/4}e^{-\vp^E/2}\). Given that the term \(RRRR\) contains four Riemann curvatures and is contracted with eight inverse metrics, one finds that for a constant dilaton, it transforms as \(e^{-2\Phi}R^E R^E R^E R^E\). Consequently, the coupling above transforms into the following expression in the Einstein frame:
\beqa
\sqrt{-\bg^E}\,e^{-\vp^E/2}R^ER^ER^ER^E\,,
\eeqa
which is invariant under S-duality.\footnote{
 The pure gravity coupling in the large-radius limit is $\sqrt{-\bg}\,e^{\vp/2}RRRR$. It transforms to $\sqrt{-\bg^E}\,e^{\vp^E/2}e^{\Phi/2}R^ER^ER^ER^E$ in the Einstein frame, which is not invariant under S-duality. However, the analogous tree-level coupling is $\sqrt{-\bg}\,e^{-2\Phi}e^{\vp/2}RRRR$, which transforms to $\sqrt{-\bg^E}\,e^{\vp^E/2}e^{-3\Phi/2}R^ER^ER^ER^E$ in the Einstein frame. The combination of these two perturbative contributions, along with non-perturbative effects, becomes invariant under the S-duality group $SL(2,\MZ)$ \cite{Green:1997tv}.}

Next, consider the coupling with coefficient \( c_{145} \) in \reef{T55}. Its structure is:
\beqa
\sqrt{-\bg}\,e^{-\vp/2}\nabla\nabla\vp\nabla\nabla\vp\nabla\nabla\vp\nabla\nabla\vp\labell{vvvv}
\eeqa
This term involves four inverse metrics. Hence, for a constant dilaton, \( \nabla\nabla\vp\nabla\nabla\vp\nabla\nabla\vp\nabla\nabla\vp \) transforms as \( e^{-2\Phi}\nabla\nabla\vp^E\nabla\nabla\vp^E\nabla\nabla\vp^E\nabla\nabla\vp^E \).
Consequently, the coupling \reef{vvvv} transforms to the following expression in the Einstein frame:
\beqa
\sqrt{-\bg^E}\,e^{-\vp^E/2}\nabla\nabla\vp^E\nabla\nabla\vp^E\nabla\nabla\vp^E\nabla\nabla\vp^E
\eeqa
which is invariant under S-duality.

Next, consider the coupling with coefficient \( c_{139} \) in \reef{T55}. Its structure is:
\beqa
\sqrt{-\bg}\,e^{-5\vp/2}\nabla W\nabla W\nabla W\nabla W\labell{WWWW}
\eeqa
This term involves six inverse metrics. Hence, for zero RR fields and a constant dilaton, \( \nabla W\nabla W\nabla W\nabla W \) transforms as \( e^{-3\Phi}\nabla W\nabla W\nabla W\nabla W \) in the Einstein frame. Therefore, the coupling \reef{WWWW} transforms to:
\beqa
\sqrt{-\bg^E}\,e^{-5\vp^E/2}e^{-2\Phi}\nabla W\nabla W\nabla W\nabla W
\eeqa
which is invariant under S-duality by the transformation rule \reef{S-dual}.

As a final example, consider the coupling with coefficient \( c_{127} \) in \reef{T55}. Its structure is:
\beqa
\sqrt{-\bg}\,e^{-7\vp/2}R WWWWWW\labell{WWWWWW}
\eeqa
This term involves one Riemann curvature and eight inverse metrics. Hence, \( R WWWWWW \) transforms as \( e^{-7\Phi/2}R^E WWWWWW \) in the Einstein frame. The coupling in the Einstein frame thus becomes:
\beqa
\sqrt{-\bg^E}\,e^{-7\vp^E/2}e^{-3\Phi}R^E WWWWWW
\eeqa
which is invariant under S-duality according to the transformation \reef{S-dual}.

Similarly, all other couplings in \reef{ttee31} are invariant under S-duality for a constant dilaton and zero RR fields.
That is,
\beqa
S_{\rm IIB}&\rightarrow S_{\rm IIB}.
\eeqa
Hence, the one-loop effective action of type IIB theory at order $\alpha'^3$ in the small-radius limit is S-duality invariant by itself. In contrast, at the large-radius limit, the tree-level, one-loop, and non-perturbative contributions must be combined to produce an S-duality invariant action at the same order \cite{Green:1997tv}. Furthermore, if one alters the radius dependence of the couplings in \reef{ttee31}, the resulting action is no longer S-duality invariant. This indicates that the radius expansion of the one-loop effective action consists of only two distinct terms: one for the large-radius limit and one for the small-radius limit, with no intermediate corrections. This conclusion can also be reached by studying the K3 reduction of the couplings in \reef{ttee31}, which we perform in the next section.

\subsection{Reduction  on K3 and duality in 5D}

    Reducing the eight-derivative couplings on K3 using the ansatz \reef{K3red} yields both eight- and four-derivative terms; we discard the former. The non-zero curvature of the K3 surface is essential here. Specifically, the integral of the Riemann-squared term gives a topological invariant, the Euler characteristic (see, e.g., \cite{Liu:2019ses}):
    \beqa
    \frac{1}{32\pi^2}\int_{K3} d^4y\sqrt{g^4}R_{\mu\nu\alpha\beta}R^{\mu\nu\alpha\beta}&=&24\,.\labell{K3int}
    \eeqa
    This fixed value is central to producing the four-derivative couplings when reducing eight-derivative terms involving $R_{\mu\nu\alpha\beta}R^{\mu\nu\alpha\beta}$.

Using the topological constraint in \reef{K3int}, the K3 reduction of the nine-dimensional one-loop gravity couplings in \reef{ttee31} produces the following four-derivative term in five dimensions:
\beqa
S_{5D}&\!\!\!\!\!=\!\!\!\!\!&-\frac{2}{\kappa^2}\frac{8\pi^5\alpha'^3}{3}\int d^{5}x\sqrt{-G^5}e^{-\vp/2}\Big[\frac{3}{2} R_{abcd} R^{abcd} -  \frac{7}{8} 
R_{abcd} W^{ab} W^{cd}e^{-\vp} -  \frac{7}{8} R_{acbd} 
W^{ab} W^{cd}e^{-\vp}\nn\\&& + \frac{7079}{48} W_{a}{}^{c} W_{bc} \nabla^{a}\vp 
\nabla^{b}\vp e^{-\vp} + \frac{3535}{24} \nabla_{a}\vp 
\nabla^{a}\vp \nabla_{b}\vp \nabla^{b}\vp + 
\frac{25111}{288} \nabla^{a}\vp \nabla_{b}\nabla_{a}\vp 
\nabla^{b}\vp \nn\\&&+ \frac{886}{3} \nabla_{b}\nabla_{a}\vp 
\nabla^{b}\nabla^{a}\vp + \frac{24031}{576} W_{a}{}^{c} W^{ab} 
\nabla_{c}\nabla_{b}\vp e^{-\vp} + \frac{9}{8} \nabla_{b}W_{ac} 
\nabla^{c}W^{ab}e^{-\vp}
\Big].\labell{SR211}
\eeqa
The resulting five-dimensional effective action is formulated in the specific minimal scheme developed in the Appendix. To analyze this action under the map in \reef{map}, we must first extend it to its most general form—a form that differs only by five-dimensional field redefinitions, integration by parts, and applications of the Bianchi identities.

We generalize the action by first constructing a maximal basis of 19 independent five-dimensional couplings (modulo total derivatives and Bianchi identities), following the procedure in the Appendix but omitting field redefinitions. Equating this basis with \reef{SR211} and subsequently incorporating field redefinitions, total derivatives, and Bianchi identities, we arrive at the following action:
\beqa
S_{5D}&=&-\frac{2}{\kappa^2}\frac{8\pi^5\alpha'^3}{3}\int d^{5}x\sqrt{-G^5}e^{-\vp/2}\Big[  \frac{3}{2} R_{abcd} 
R^{abcd} + a_{4} R W_{ab} W^{ab}e^{-\vp} + a_{5} R^{ab} 
W_{a}{}^{c} W_{bc}e^{-\vp} \nn\\&&+a_1 R_{ab} R^{ab}- 2 (3 + a_1 + 2 a_{10} - 2 a_{12} - 4 
a_{14} + 4 a_{16} + 8 a_{18}) R^2\nn\\&& + \frac{1}{16} (3 + 2 a_1 - 4 a_{12} - 8 a_{14} + 8 a_{16} 
+ 16 a_{17}) W_{a}{}^{c} W^{ab} W_{b}{}^{d} W_{cd} e^{-2\vp}\nn\\&&+ \frac{1}{8} 
\bigl(-3 a_1 + 2 (-6 + a_{12} + 2 a_{14} - 2 a_{16} - 4 a_{17} - 2 a_{5})\bigr) 
R_{abcd} W^{ab} W^{cd}e^{-\vp} \nn\\&&+ a_8 W_{ab} W^{ab} W_{cd} W^{cd}e^{-2\vp} + 
a_{10} R \nabla_{a}\vp \nabla^{a}\vp + a_{11} W_{bc} W^{bc} 
\nabla_{a}\vp \nabla^{a}\vp e^{-\vp}\nn\\&&+ a_{16} R_{ab} 
\nabla^{a}\vp \nabla^{b}\vp + a_{17} W_{a}{}^{c} W_{bc} 
\nabla^{a}\vp \nabla^{b}\vp e^{-\vp}+ a_{18} \nabla_{a}\vp 
\nabla^{a}\vp \nabla_{b}\vp \nabla^{b}\vp \nn\\&&-  a_{19} \nabla^{a}
\vp \nabla_{b}\nabla_{a}\vp \nabla^{b}\vp -  a_{12} 
R_{ab} \nabla^{b}\nabla^{a}\vp -  a_{13} W_{a}{}^{c} W_{bc} 
\nabla^{b}\nabla^{a}\vp e^{-\vp} \nn\\&&+ a_{14} \nabla_{b}\nabla_{a}\vp 
\nabla^{b}\nabla^{a}\vp+ a_{15} W^{bc} \nabla^{a}\vp 
\nabla_{c}W_{ab} e^{-\vp}\nn\\&&+ \frac{1}{8} (6 + 3 a_1 - 2 a_{12} - 4 a_{14} + 4 a_{16} + 
8 a_{17} + 4 a_{5}) \nabla_{c}W_{ab} \nabla^{c}W^{ab}e^{-\vp}
\Big],\labell{SR22}
\eeqa
 with \( a_{1}, a_{4}, a_{5}, \cdots \) being 14 arbitrary scheme parameters. The actions \reef{SR22} and \reef{SR211} are physically equivalent, related by field redefinitions.

Under the map in \reef{map}, the couplings above produce the following dual action:
\beqa
S^{\rm dual}_{5D}&=&-\frac{2}{\kappa^2}\frac{8\pi^5\alpha'^3}{3}\int d^{5}x\sqrt{-G^5}e^{-2\Phi}\Big[ \frac{3}{2} R_{abcd} 
R^{abcd} + a_{4} R V_{ab}V^{ab} + a_{5} R^{ab} 
V_{a}{}^{c} V_{bc}+\cdots
\Big],\labell{SR220}
\eeqa
where the ellipsis represents other lengthy couplings not written explicitly.

In contrast, the ten-dimensional heterotic theory at tree-level contains four-derivative gravity-dilaton couplings of the form \cite{Metsaev:1987zx}:
\beqa
S^H_{R^2}&=&-\frac{2}{\kappa^2_{10}}\frac{\alpha'}{8}\int d^{10}x\sqrt{-G}e^{-2\Phi}R_{\mu\nu\alpha\beta}R^{\mu\nu\alpha\beta}\,,\labell{SHet}
\eeqa
where $\kappa^2_{10}=\kappa^2 g_s^2$. The dimensional reduction of this term on a five-torus $T^5$ of volume $V$, considering only one non-zero vector arising from the metric on one circle of the $T^5$, yields the following five-dimensional gravity-dilaton-vector couplings \cite{Garousi:2025wfk}:
\beqa
S^H_{R^2}&=&-\frac{2}{\kappa^2_{10}}\frac{\alpha'V}{12}\int d^{5}x\sqrt{-G^5}e^{-2\Phi}\Big[\frac{3}{2} R_{abcd} R^{abcd} + \frac{15}{16} 
V_{a}{}^{c} V^{ab} V_{b}{}^{d} V_{cd} -  \frac{3}{2} 
R_{abcd} V^{ab} V^{cd} \nn\\&&-  \frac{3}{2} R_{acbd} 
V^{ab} V^{cd} + \frac{9}{16} V_{ab} V^{ab} V_{cd} V^{cd} + 
\frac{3}{2} \nabla_{c}V_{ab} \nabla^{c}V^{ab}\Big]\,.\labell{S'6}
\eeqa
Using the relation in \reef{NS5F}, one finds this action is identical to the action in \reef{SR220} after an appropriate field redefinition, provided the following relations hold between the parameters in \reef{SR220}:
\beqa
&&a_{18} = 
  9/4 + a_{1}/2 - a_{10} + 4 a_{11} + a_{12}/8 - 4 a_{13} + (9 a_{14})/4 - 
   2 a_{15} - (5 a_{16})/4 - a_{17}, \nn\\&&
 a_{19} = 6 + (3 a_{1})/2 - 2 a_{10} + 8 a_{11} - a_{12}/4 - 8 a_{13} + 
   5 a_{14} - 4 a_{15} - (3 a_{16})/2 - 2 a_{17}, \nn\\&&
 a_{5} = -(3/4) - (5 a_{1})/8 + (3 a_{12})/4 - 
   2 a_{13} + (3 a_{14})/2 - (3 a_{16})/2 - 3 a_{17}, \nn\\&&
 a_{8} = 381/128 + (135 a_{1})/256 + a_{4}/4 - (3 a_{10})/4 + 
   4 a_{11} - (5 a_{12})/128 \nn\\&&\qquad\quad- (31 a_{13})/8 + (115 a_{14})/
    64 - (31 a_{15})/16 - (55 a_{16})/64 - (39 a_{17})/32
\eeqa
Substituting these relations into \reef{SR22}, we find that the five-dimensional action \reef{SR22}—with its 10 arbitrary parameters—is dual to the heterotic couplings at order $\alpha'$ on $T^5$. It is interesting to note that this result is analogous to the K3 reduction of type IIA theory in the metric-dilaton-RR one-form sector \cite{Garousi:2025wfk}, which also features 10 arbitrary parameters and is dual to the heterotic couplings at order $\alpha'$ on $T^4$.

    Since the couplings in the heterotic theory at order $\alpha'$ are exact \cite{Ellis:1987dc,Ellis:1989fi}, the duality \reef{map}—which relates the radius parameter $\varphi$ of type IIB to the dilaton of the heterotic theory—confirms the S-duality result that the radius expansion of the couplings is truncated, consisting only of a small-radius limit and a large-radius limit. Note that, upon reduction on K3, the large-radius limit produced by the KK reduction of the couplings in \reef{IIB} does not produce couplings at order $\alpha'$.

\section{Conclusion}

It is known in the literature that the S-duality of the 10-dimensional type IIB effective action at order $\alpha'^3$ involves contributions from the perturbative sphere- and torus-levels, as well as non-perturbative effects, which combine to achieve $SL(2,\mathbb{Z})$ invariance \cite{Green:1997tv}. The circular reduction of these combinations also results in invariant terms in nine dimensions \cite{Green:1997di}.

In this paper, we have shown that purely stringy couplings, which arise at loop level from winding modes on the circle, correspond to inherently nine-dimensional terms in the effective action. These couplings are not descended from ten-dimensional terms and are separately invariant under the continuous \( SL(2,\mathbb{R}) \) symmetry. In particular, we have derived two distinct classes of such terms: the inherently nine-dimensional Chern–Simons coupling, obtained by applying T-duality to the KK reduction of the type IIA Chern–Simons term; and the inherently nine-dimensional couplings in the metric–radion–vector sector of type IIB theory, obtained by imposing T-duality on the KK reduction of the pure gravity couplings of type IIA theory. Both sets of couplings are invariant under \( SL(2,\mathbb{R}) \) transformations when the dilaton is held constant and the RR fields are set to zero.

We have seen that S-duality dictates the radius-dependence of the nine-dimensional couplings. This dependence corresponds to one of two limits: the large-radius limit, which is associated with the KK reduction of the 10-dimensional type IIB couplings, or the small-radius limit, which is associated with the T-dual of the KK reduction of the 10-dimensional type IIA couplings.

        We have also investigated the reduction of the resulting nine-dimensional type IIB couplings at order $\alpha'^3$ on a K3 surface. While the large-radius limit yields no $\alpha'$-order couplings, the small-radius limit does produce them. The resulting couplings are consistent with the four-derivative effective action of the heterotic theory on $T^5$ after appropriate field redefinitions. Crucially, in this field redefinition, the radius of the circle in type IIB theory maps to the dilaton of the heterotic theory. Given that the dilaton dependence of the four-derivative couplings in the heterotic theory is exact  \cite{Ellis:1987dc,Ellis:1989fi}, this equivalence strongly confirms that the nine-dimensional coupling constant receives contributions exclusively from the large-radius and small-radius limits.

 The symmetry structure of the eight-derivative couplings can be understood through the duality between M-theory on \( T^{2} \) and type IIB on \( S^{1} \) \cite{Schwarz:1996qw,Becker:2007zj}. Our analysis shows that the \( \alpha'^3 \) couplings in type IIB are \( SL(2,\mathbb{Z}) \)-invariant at large radius and \( SL(2,\mathbb{R}) \)-invariant at small radius. Consequently, the effective action of M-theory on \( T^{2} \) must inherit this behavior: it is invariant under the discrete \( SL(2,\mathbb{Z}) \) for a large type IIB circle and under the continuous \( SL(2,\mathbb{R}) \) for a small circle.  This observation was made in \cite{Green:1997as} for the nine-dimensional \(R^4\) couplings. The classical \( SL(2,\mathbb{R}) \) symmetry of the above \(\alpha'^3\) coupling indicates that these couplings should be reproduced by one-loop amplitudes derived from ten-dimensional type IIB supergravity compactified on a circle, following the same approach as in \cite{Green:1997as}.

We demonstrate S-duality for the one-loop couplings in nine-dimensional type IIB theory at small radius, under the assumption of a constant dilaton and vanishing RR fields. The first step is to extend this result by incorporating RR fields and a non-constant dilaton into the nine-dimensional type IIB framework, maintaining S-duality invariance. We then apply T-duality to this generalized action to derive the corresponding one-loop couplings in nine-dimensional type IIA theory at large radius. Lifting these results to ten dimensions yields the one-loop $\alpha'^3$ terms in type IIA theory. These terms should originate from the circle reduction of eleven-dimensional M-theory couplings, which relates them directly to M-theory. In line with this expectation, recent results in \cite{Garousi:2025dil,Garousi:2025wfk} have identified related couplings involving RR fields and the dilaton that match the M-theory Chern-Simons term and pure gravity couplings. A compelling future check would be to explicitly show that the circular reduction of the type IIA couplings found in these references indeed exhibits the expected \(SL(2,\mathbb{R})\) symmetry in nine dimensions after a T-duality transformation.

Using the fact that dimensional reduction of M-theory covariant couplings on \(S^1 \times S^1\) produces Lorentz-invariant couplings in nine-dimensional type IIA theory with a global \(O(2)\) symmetry \cite{Bergshoeff:1995as}, one expects that any eleventh-dimensional couplings at the eighth-derivative order would yield corresponding couplings in type IIA theory possessing the same \(O(2)\) symmetry. After applying T-duality, these would produce couplings in type IIB theory invariant under \(SL(2,\mathbb{R})\) transformations. Consequently, this nine-dimensional symmetry alone cannot serve as a constraint to uniquely determine the M-theory couplings.

We have seen that the nine-dimensional one-loop couplings in type IIB theory at order \(\alpha'^3\) exhibit S-duality symmetry as \(SL(2,\mathbb{R})\) in the small-radius limit, and as \(SL(2,\mathbb{Z})\) in the large-radius limit. This indicates that the S-duality invariant couplings in the large-radius limit include tree-level, one-loop, and non-perturbative effects, whereas in the small-radius limit there are no couplings at order \(\alpha'^3\) other than the one-loop contribution. These observations are consistent with supersymmetry, which implies that there are only two classes of supersymmetrizable \(R^4\) corrections in nine dimensions \cite{Green:1998by}, and that the \(R^4\) couplings receive no contributions from two-loop or higher loops \cite{Green:1998by,Berkovits:1997pj,Basu:2011he,Bossard:2014lra,DHoker:2005vch}.



\newpage
\vskip 0.5 cm
{\Large \bf Appendix: }{\large\bf Minimal basis in the metric-radion-vector Sector}
\vskip 0.5 cm

 This appendix constructs the minimal basis of eight-derivative couplings for the metric-radion-vector sector. We first generate all covariant, \( U(1) \)-invariant terms with an even number of field strengths \( V_{ab} \), ensuring the correct radion dependence. Our convention assigns a factor of \( e^{\varphi/2} \) to each \( V \), with an additional overall factor of \( e^{\varphi/2} \).
 Using the xAct package \cite{Nutma:2013zea}, this process yields 18,462 candidate couplings, schematically represented as:
\beqa
\mathcal{L}'=& c'_1 R_{abcd}R^{abcd} \
R_{efgh}V^{ef}V^{gh}e^{3\varphi/2}+\cdots\,,\labell{L3}
\eeqa
where \( c'_1,\cdots, c'_{18462} \) are provisional coupling constants.
However, these terms are not independent, as they are subject to redundancies from: (i) total derivatives, (ii) field redefinition equivalences, and (iii) Bianchi identity constraints. A systematic reduction is therefore necessary to isolate a minimal, independent set.

To systematically eliminate redundant terms arising from total derivative ambiguities in the Lagrangian \(\mathcal{L}'\), we introduce the following compensating terms to the action:
\beqa
-\frac{4\pi}{\kappa^2}\int d^9x\sqrt{-\bg}\,\mathcal{J}&\equiv& -\frac{4\pi}{\kappa^2}\int d^9x\sqrt{-\bg} \,\nabla_a (e^{\varphi/2}{\cal I}^a)\,.\labell{J3}
\eeqa
Here, the vector \({\cal I}^a\) encompasses all possible covariant and gauge-invariant seven-derivative terms constructed from the metric, radius, and gauge fields. Each gauge field strength must be accompanied by a factor of \(e^{\varphi/2}\). Our systematic classification identifies 10,349 such independent vectors, with corresponding coefficients \(J_1,\cdots, J_{10349}\).

To eliminate redundant terms from field redefinition ambiguities in \(\mathcal{L}'\), we begin with the KK reduction of the ten-dimensional tree-level coupling \(e^{-2\Phi}R\) on a circle. Using the metric reduction in \reef{reduc}, this yields the following nine-dimensional action:
\beqa
S_0&=&-\frac{4\pi}{\kappa^2}\int d^9x\sqrt{-\bg}e^{-2\Phi+\vp/2}\Big[R-\frac{1}{4}\nabla_a\vp\nabla^a\vp-\frac{e^{\vp}}{4}V_{ab}V^{ab}\Big].
\eeqa
Up to total derivative terms, the variation of this action for a constant dilaton yields:
\beqa
\delta S_0&=&-\frac{4\pi}{\kappa^2}\int d^9x\sqrt{-\bg}e^{-2\Phi+\vp/2}\Big[D\Phi\,\delta\Phi+D\vp\,\delta\vp+Dg^a\,\delta g_a+D\bg^{ab}\,\delta\bg_{ab}\Big]\nn\\&\equiv&-\frac{4\pi}{\kappa^2}\int d^9x\sqrt{-\bg}\,\mathcal{K},
\eeqa
where
\beqa
D\Phi&=&-2 R + \frac{1}{2} e^{ \vp} V_{ab} V^{ab} + \frac{1}{2} \nabla_{a}\vp \nabla^{a}\vp\nn\\
D\vp&=&\frac{1}{2} R - \frac{3}{8} e^{ \vp} V_{ab} V^{ab} + \frac{1}{2} \nabla_{a}\nabla^{a}\vp + \frac{1}{8} \nabla_{a}\vp \nabla^{a}\vp\nn\\
Dg^a&=&- e^{ \vp} \nabla_{b}V^{ab} - \frac{3}{2} e^{ \vp} V^{a}{}_{b} \nabla^{b}\vp\nn\\
D\bg^{ab}&=&- R^{ab} + \frac{1}{2} G^{ab} R + \frac{1}{2} e^{ \vp} V^{ac} V^{b}{}_{c} - \frac{1}{8} e^{ \vp} G^{ab} V_{cd} V^{cd} \nn\\&&+ \frac{1}{2} \nabla^{a}\vp \nabla^{b}\vp + \frac{1}{2} \nabla^{b}\nabla^{a}\vp - \frac{1}{2} G^{ab} \nabla_{c}\nabla^{c}\vp - \frac{3}{8} G^{ab} \nabla_{c}\vp \nabla^{c}\vp.
\eeqa
It is important to note that although the dilaton is constant in the background configuration, its variations under field redefinition are non-zero and must be included.

The field redefinition contributions arise from the following infinitesimal transformations of the fundamental fields:
\begin{eqnarray}
\bg_{ab}&\rightarrow &\bg_{ab}+\alpha'^3 e^{2\Phi}\delta\bg_{ab},\nn\\
g_{a}&\rightarrow &g_{a}+ \alpha'^3 e^{2\Phi-\vp/2}\delta g_{a},\nn\\
\vp &\rightarrow &\vp+\alpha'^3 e^{2\Phi}\delta\vp,\nn\\
\Phi &\rightarrow &\Phi+\alpha'^3 e^{2\Phi}\delta\Phi,\labell{gbp}
\end{eqnarray}
where \(\delta g_{a}\) contains odd powers of the \(U(1)\) gauge field strength, while \(\delta \bg_{ab}\), \(\delta\vp\), and \(\delta\Phi\) contain even powers. Specifically, this manifests as 3,265 metric perturbations (coefficients \(\{e_i\}\)), 1,621 gauge-field perturbations \(\{g_i\}\), 656 radius perturbations \(\{f_i\}\), and 656 dilaton perturbations \(\{h_i\}\).

By augmenting \(\mathcal{L}'\) with these field redefinitions and the previously introduced total derivative terms, we obtain an equivalent Lagrangian \(\mathcal{L}\) with transformed parameters \(\{c_i\}\), yielding the fundamental constraint:
\beqa
\Delta - \mathcal{J} - \mathcal{K} = 0.\labell{DLK}
\eeqa
The difference \(\Delta = \mathcal{L} - \mathcal{L}'\) preserves the same functional form as \(\mathcal{L}'\) but with modified coefficients \(\delta c_i = c_i - c'_i\), representing the net effect of the field redefinitions and total derivative terms.

To systematically solve equation \reef{DLK}, we must first express it in terms of linearly independent couplings by enforcing the relevant Bianchi identities:
\beqa
 R_{\alpha[\beta\gamma\delta]}&=&0\,,\nn\\
 \nabla_{[\mu}R_{\alpha\beta]\gamma\delta}&=&0\,,\labell{bian}\\
\nabla_{[\mu}V_{\alpha\beta]}&=&0\,,\nn\\
{[}\nabla,\nabla{]}\mathcal{O}-R\mathcal{O}&=&0\,.\nn
\eeqa
 Our procedure for implementing the Bianchi identities in a non-gauge-invariant formulation involves two key steps: (1) adopting a local inertial frame to simplify derivatives, and (2) expressing all \(\partial V\) terms in \reef{DLK} through the relation \(V = dg\). As shown in \cite{Garousi:2019cdn}, this combined approach automatically satisfies all Bianchi identities for the curvature and $U(1)$ field strength.

Through this systematic procedure, all terms on the left-hand side of \reef{DLK} can be expressed as a linear combination of independent (though non-gauge-invariant) couplings. Setting their coefficients to zero yields a system of algebraic equations with two distinct classes of solutions: (i) 298 relations involving only the variations \(\delta c_i\), and (ii) additional equations that mix \(\delta c_i\) with the coefficients of total derivative and field redefinition terms, which we disregard.

The invariant count of 298 relations in the first class determines the dimension of the physically meaningful coupling space in \({\cal L}'\). This number remains unchanged under scheme transformations; while a particular scheme may nullify certain coefficients in \({\cal L}'\), substituting these conditions back into \reef{DLK} preserves exactly 298 constraints among the \(\delta c_i\) parameters.

To systematically eliminate redundant couplings while preserving the 298 fundamental relations among the \(\delta c_i\) parameters, we impose a specific scheme choice by nullifying all coefficients in \({\cal L}'\) that contain the following structures: \(R\), \(\nabla_a V^{ab}\), \(\nabla_a\nabla^a\vp\), \(\nabla_a\nabla_b V_{cd}\), \(\nabla_a\nabla_b \nabla_{c}\vp\), \(\nabla_a R_{bcde}\), \(R_{ab}R_{cd}\), \(R_{abcd}R_{ef}\), or \(V_{ab}R_{cd}\). Crucially, this elimination preserves exactly 298 constraints among the remaining \(\delta c_i\) variations, demonstrating that the eliminated terms represent non-essential redundancies in the effective action.

We next eliminate all couplings in \({\cal L}'\) containing \(R_{ab}\) by setting their coefficients to zero. Solving \reef{DLK} under this constraint yields 297 relations among the \(\delta c_i\) parameters, demonstrating that at least one independent coupling must contain Ricci curvature. To identify this essential coupling, we implemented a binary search strategy: we divided the Ricci-containing terms into subsets, nullified one subset's coefficients, and verified whether \reef{DLK} generated 298 relations among the remaining \(\delta c_i\) parameters. If not, we retained the complementary subset. Through iterative application of this method, we uniquely identified the independent coupling:
\beqa
[R^2 R'']_1 &=& c_{268}  R^{abcd} R^{ef}{}_{cd} \nabla_{a}
\nabla_{e}R_{bf}\,.\labell{T53}
\eeqa
This independent coupling also appears in the minimal basis for metric-dilaton-RR one-form couplings at order \(\alpha'^3\) \cite{Garousi:2025wfk}.

 Although numerous coefficient choices satisfy the 298 relations \(\delta c_i = 0\), our chosen scheme organizes the couplings into 49 distinct structures. These fundamental structures, which form the basis, are explicitly listed below:
\beqa
{\cal L}(\vp,V)&\!\!\!\!=\!\!\!\! &e^{\vp/2}\Big([R^2R'']_1+[R^4]_7+[\vp'^4]_2+[V'^4]_{5}+[RV'^2\vp'^2]_{11}+[RV'^2\vp'']_{9}+[R^2V'^2]_{13}\nn\\&&+[R^3V'^2]_{14}+
[R^3\vp'']_3+[R^3\vp'^2]_5+[RV^2\vp'^2\vp'']_7+[R^2\vp''^2]_5+[R^2\vp'^2\vp'']_4\nn\\&&+[R^2\vp'^4]_3+[V^3V'\vp'^3]_{5}+[R^2V^4]_{13}+[R^2V^2\vp'']_{13}+[R^2V^2\vp'^2]_{17}+[RV^4\vp'^2]_{9}\nn\\&&+[RV^2V'^2]_{27}+[RV^6]_1+[V^4\vp'^2\vp'']_1+[R^2VV'\vp']_{14}+[V^4\vp'^4]_{2}+[RV^2\vp'^4]_4\nn\\&&+[RVV'\vp'^3]_{10}+[RV^4\vp'']_{7}+[V^2\vp'^6]_1+[V^4\vp''^2]_5+[RV^2\vp''^2]_8+[V^2V'^2\vp'']_{15}\nn\\&&+[RV^3V'\vp']_{1}+[RVV'\vp'\vp'']_2+[V^2\vp'^2\vp''^2]_2+[V^2V'^2\vp'^2]_{11}+[VV'\vp'\vp''^2]_{1}\labell{T55}\\&&+[VV'\vp'^3\vp'']_4+[R\vp'^2\vp''^2]_2+[V'^2\vp'^4]_4+[VV'\vp'^5]_1+[V^2\vp'^4\vp'']_2+[R\vp''^3]_1\nn\\&&+[R\vp'^4\vp'']_1+[V'^2\vp''^2]_6+[V'^2\vp'^2\vp'']_4+[VV'^3\vp']_1+[\vp'^4\vp''^2]_2+[\vp'^8]_1+[V^4V'^2]_{11}
\Big)\,.\nn
\eeqa
The notation $[X]_n$ denotes that structure $X$ admits $n$ distinct contractions, each with an independent coupling constant. A prime symbol  indicates covariant differentiation of the associated field. Among these structures, the coupling $[R^2R'']_1$ is explicitly given in \reef{T53}, while all others are the following:
\beqa
[R^4]_7&=&c_{117}  R^{abcd} R_{c}{}^{e}{}_{a}{}^{f} 
R_{d}{}^{g}{}_{b}{}^{h} R_{fheg} +  
 c_{253}  R_{a}{}^{efg} R^{abcd} 
R_{c}{}^{h}{}_{bf} R_{ghde} +  
 c_{254}  R_{a}{}^{e}{}_{c}{}^{f} R^{abcd} 
R_{be}{}^{gh} R_{ghdf}\nn\\&& +  
 c_{255}  R_{acb}{}^{e} R^{abcd} 
R_{d}{}^{fgh} R_{ghef} +  
 c_{256}  R^{abcd} R_{cdab} 
R^{efgh} R_{ghef} +  
 c_{266}  R_{ab}{}^{ef} R^{abcd} 
R_{ghef} R^{gh}{}_{cd} \nn\\&&+  
 c_{267}  R_{ab}{}^{ef} R^{abcd} 
R_{ghdf} R^{gh}{}_{ce} \,,
\eeqa
\beqa
[\vp'^4]_2&=&c_{145}  \nabla_{a}\nabla^{c}\vp  
\nabla^{a}\nabla^{b}\vp  \nabla_{b}\nabla^{d}\vp  \nabla_{d}
\nabla_{c}\vp  +  
 c_{147}  \nabla^{a}\nabla^{b}\vp  \nabla_{b}\nabla_{a}
\vp  \nabla^{c}\nabla^{d}\vp  \nabla_{d}\nabla_{c}\vp
 \,,
\eeqa
\beqa
[V'^4]_{5}&=&e^{2 \vp }  
 c_{139}  \nabla^{a}V^{bc} \nabla_{b}V_{a}{}^{d} 
\nabla_{c}V^{ef} \nabla_{d}V_{ef} + e^{2 \vp }  
 c_{196}  \nabla_{a}V_{bc} \nabla^{a}V^{bc} 
\nabla_{d}V_{ef} \nabla^{d}V^{ef}\nn\\&& + e^{2 \vp }  
 c_{216}  \nabla^{a}V^{bc} \nabla_{b}V_{c}{}^{d} 
\nabla_{d}V^{ef} \nabla_{e}V_{af} + e^{2 \vp }  
 c_{249}  \nabla^{a}V^{bc} \nabla_{b}V_{a}{}^{d} 
\nabla^{e}V_{c}{}^{f} \nabla_{f}V_{de}\nn\\&& + e^{2 \vp }  
 c_{250}  \nabla_{a}V^{de} \nabla^{a}V^{bc} 
\nabla_{f}V_{ce} \nabla^{f}V_{bd}\,,
\eeqa
\beqa
[RV'^2\vp'^2]_{11}&=&e^{\vp }  
 c_{91}  R_{efbd} \nabla_{a}\vp  \nabla^{a}\vp 
 \nabla^{b}V^{cd} \nabla_{c}V^{ef} + e^{\vp }  
 c_{138}  R_{aebf} \nabla^{a}\vp  
\nabla^{b}\vp  \nabla^{c}V^{de} \nabla_{d}V_{c}{}^{f}\nn\\&& + e^{\vp 
} 
  c_{140}  R_{afbc} \nabla^{a}\vp  \nabla^{b}\vp  
\nabla^{c}V^{de} \nabla_{d}V_{e}{}^{f} + e^{\vp }  
 c_{194}  R_{efcd} \nabla_{a}V_{b}{}^{c} 
\nabla^{a}\vp  \nabla^{b}\vp  \nabla^{d}V^{ef} \nn\\&&+ e^{\vp } 
  c_{233}  R_{efcd} \nabla_{a}V^{cd} 
\nabla^{a}\vp  \nabla^{b}\vp  \nabla^{e}V_{b}{}^{f} + e^{\vp 
} 
  c_{236}  R_{dfce} \nabla^{a}\vp  \nabla^{b}\vp  
\nabla^{c}V_{a}{}^{d} \nabla^{e}V_{b}{}^{f} \nn\\&&+ e^{\vp }  
 c_{238}  R_{dfbe} \nabla_{a}\vp  
\nabla^{a}\vp  \nabla^{b}V^{cd} \nabla^{e}V_{c}{}^{f} + e^{\vp 
} 
  c_{239}  R_{dfbe} \nabla_{a}V^{cd} \nabla^{a}\vp  
\nabla^{b}\vp  \nabla^{e}V_{c}{}^{f} \nn\\&&+ e^{\vp }  
 c_{240}  R_{efbd} \nabla_{a}V^{cd} 
\nabla^{a}\vp  \nabla^{b}\vp  \nabla^{e}V_{c}{}^{f} + e^{\vp 
} 
  c_{242}  R_{dfbe} \nabla^{a}\vp  \nabla^{b}\vp  
\nabla^{c}V_{a}{}^{d} \nabla^{e}V_{c}{}^{f} \nn\\&&+ e^{\vp }  
 c_{244}  R_{efbd} \nabla^{a}\vp  
\nabla^{b}\vp  \nabla^{c}V_{a}{}^{d} \nabla^{e}V_{c}{}^{f}\,,\labell{RFFP}
\eeqa
\beqa
[RV'^2\vp'']_{9}&=&e^{\vp }  
 c_{149}  R_{efac} \nabla^{a}V^{bc} 
\nabla_{b}V^{de} \nabla_{d}\nabla^{f}\vp  + e^{\vp }  
 c_{172}  R_{dfce} \nabla_{a}\nabla^{f}\vp  
\nabla^{a}V^{bc} \nabla^{d}V_{b}{}^{e} \nn\\&&+ e^{\vp }  
 c_{173}  R_{efcd} \nabla_{a}\nabla^{f}\vp  
\nabla^{a}V^{bc} \nabla^{d}V_{b}{}^{e} + e^{\vp }  
 c_{176}  R_{efad} \nabla^{a}V^{bc} 
\nabla_{c}\nabla^{f}\vp  \nabla^{d}V_{b}{}^{e} \nn\\&&+ e^{\vp }  
 c_{193}  R_{efcd} \nabla^{a}V^{bc} 
\nabla_{b}\nabla_{a}\vp  \nabla^{d}V^{ef} + e^{\vp }  
 c_{218}  R_{dfbc} \nabla^{a}V^{bc} 
\nabla^{d}V^{ef} \nabla_{e}\nabla_{a}\vp \nn\\&& + e^{\vp }  
 c_{220}  R_{cfad} \nabla^{a}V^{bc} 
\nabla^{d}V^{ef} \nabla_{e}\nabla_{b}\vp  + e^{\vp }  
 c_{247}  R_{cedf} \nabla^{a}V^{bc} 
\nabla_{b}V_{a}{}^{d} \nabla^{e}\nabla^{f}\vp \nn\\&& + e^{\vp }  
 c_{248}  R_{deaf} \nabla^{a}V^{bc} 
\nabla_{b}V_{c}{}^{d} \nabla^{e}\nabla^{f}\vp \,,\labell{RRF'F'}
\eeqa
\beqa
[R^2V'^2]_{13}&=&e^{\vp }  
 c_{271}  R_{c}{}^{efg} R_{fgde} 
\nabla_{a}V_{b}{}^{d} \nabla^{a}V^{bc} + e^{\vp }  
 c_{277}  R_{fgde} R^{fg}{}_{bc} 
\nabla_{a}V^{de} \nabla^{a}V^{bc}\nn\\&& + e^{\vp }  
 c_{278}  R_{fgce} R^{fg}{}_{bd} 
\nabla_{a}V^{de} \nabla^{a}V^{bc} + e^{\vp }  
 c_{20}  R^{defg} R_{fgde} 
\nabla^{a}V^{bc} \nabla_{b}V_{ac} \nn\\&&+ e^{\vp }  
 c_{24}  R_{c}{}^{efg} R_{fgde} 
\nabla^{a}V^{bc} \nabla_{b}V_{a}{}^{d} + e^{\vp }  
 c_{31}  R_{d}{}^{f}{}_{a}{}^{g} R_{egcf} 
\nabla^{a}V^{bc} \nabla_{b}V^{de} \nn\\&&+ e^{\vp }  
 c_{164}  R_{a}{}^{f}{}_{d}{}^{g} R_{egcf} 
\nabla^{a}V^{bc} \nabla^{d}V_{b}{}^{e} + e^{\vp }  
 c_{166}  R_{c}{}^{f}{}_{a}{}^{g} R_{egdf} 
\nabla^{a}V^{bc} \nabla^{d}V_{b}{}^{e} \nn\\&&+ e^{\vp }  
 c_{169}  R_{fgcd} R^{fg}{}_{ae} 
\nabla^{a}V^{bc} \nabla^{d}V_{b}{}^{e} + e^{\vp }  
 c_{186}  R_{b}{}^{g}{}_{ac} R_{egdf} 
\nabla^{a}V^{bc} \nabla^{d}V^{ef} \nn\\&&+ e^{\vp }  
 c_{188}  R_{a}{}^{g}{}_{de} R_{fgbc} 
\nabla^{a}V^{bc} \nabla^{d}V^{ef} + e^{\vp }  
 c_{191}  R_{bea}{}^{g} R_{fgcd} 
\nabla^{a}V^{bc} \nabla^{d}V^{ef} \nn\\&&+ e^{\vp }  
 c_{192}  R_{b}{}^{g}{}_{ae} R_{fgcd} 
\nabla^{a}V^{bc} \nabla^{d}V^{ef}\,,
\eeqa
\beqa
[R^3V'^2]_{14}&=&e^{\vp }  
 c_{109}  R_{d}{}^{g}{}_{c}{}^{h} 
R^{de}{}_{b}{}^{f} R_{fheg} V_{a}{}^{c} V^{ab} + 
e^{\vp }  
 c_{243}  R_{b}{}^{d}{}_{c}{}^{e} 
R_{d}{}^{fgh} R_{ghef} V_{a}{}^{c} V^{ab} \nn\\&&+ 
e^{\vp }  
 c_{264}  R_{b}{}^{def} R_{ghdf} 
R^{gh}{}_{ce} V_{a}{}^{c} V^{ab} + e^{\vp }  
 c_{85}  R_{a}{}^{e}{}_{c}{}^{f} 
R_{e}{}^{g}{}_{b}{}^{h} R_{fhdg} V^{ab} V^{cd}\nn\\&& + 
e^{\vp }  
 c_{99}  R_{a}{}^{e}{}_{c}{}^{f} 
R_{d}{}^{g}{}_{b}{}^{h} R_{fheg} V^{ab} V^{cd} + 
e^{\vp }  
 c_{202}  R_{a}{}^{efg} R_{f}{}^{h}{}_{be} 
R_{ghcd} V^{ab} V^{cd} \nn\\&&+ e^{\vp }  
 c_{213}  R_{a}{}^{efg} R_{f}{}^{h}{}_{bc} 
R_{ghde} V^{ab} V^{cd} + e^{\vp }  
 c_{234}  R_{a}{}^{e}{}_{bc} R_{d}{}^{fgh} 
R_{ghef} V^{ab} V^{cd} \nn\\&&+ e^{\vp }  
 c_{252}  R_{acbd} R^{efgh} 
R_{ghef} V^{ab} V^{cd} + e^{\vp }  
 c_{260}  R_{a}{}^{e}{}_{c}{}^{f} R_{ghef} 
R^{gh}{}_{bd} V^{ab} V^{cd} \nn\\&&+ e^{\vp }  
 c_{261}  R^{ef}{}_{ac} R_{ghdf} 
R^{gh}{}_{be} V^{ab} V^{cd} + e^{\vp }  
 c_{262}  R_{a}{}^{e}{}_{c}{}^{f} R_{ghde} 
R^{gh}{}_{bf} V^{ab} V^{cd}\nn\\&& + e^{\vp }  
 c_{263}  R_{a}{}^{e}{}_{b}{}^{f} R_{ghef} 
R^{gh}{}_{cd} V^{ab} V^{cd} + e^{\vp }  
 c_{265}  R^{ef}{}_{ab} R_{ghdf} 
R^{gh}{}_{ce} V^{ab} V^{cd} \,,
\eeqa
\beqa
[R^3\vp'']_3&=&c_{297}  R_{c}{}^{f}{}_{b}{}^{g} 
R^{cd}{}_{a}{}^{e} R_{egdf} 
\nabla^{a}\nabla^{b}\vp  +  
 c_{7}  R_{a}{}^{c}{}_{b}{}^{d} 
R_{c}{}^{efg} R_{fgde} \nabla^{a}\nabla^{b}\vp  
\nn\\&&+ 
  c_{13}  R_{a}{}^{cde} R_{fgce} 
R^{fg}{}_{bd} \nabla^{a}\nabla^{b}\vp \,,
\eeqa
\beqa
[R^3\vp'^2]_5&=&c_{282}  R^{bcde} R_{d}{}^{f}{}_{b}{}^{g} 
R_{egcf} \nabla_{a}\vp  \nabla^{a}\vp  +  
 c_{283}  R_{bd}{}^{fg} R^{bcde} 
R_{fgce} \nabla_{a}\vp  \nabla^{a}\vp \nn\\&& +  
 c_{65}  R_{c}{}^{f}{}_{b}{}^{g} 
R^{cd}{}_{a}{}^{e} R_{egdf} \nabla^{a}\vp  
\nabla^{b}\vp  +  
 c_{74}  R_{a}{}^{c}{}_{b}{}^{d} 
R_{c}{}^{efg} R_{fgde} \nabla^{a}\vp  
\nabla^{b}\vp  \nn\\&&+  
 c_{80}  R_{a}{}^{cde} R_{fgce} 
R^{fg}{}_{bd} \nabla^{a}\vp  \nabla^{b}\vp,
\eeqa
\beqa
[RV^2\vp'^2\vp'']_7&=&e^{\vp }  
 c_{81}  R_{cebf} V_{d}{}^{f} V^{de} 
\nabla_{a}\nabla^{c}\vp  \nabla^{a}\vp  \nabla^{b}\vp  + 
e^{\vp }  
 c_{83}  R_{efcd} V_{b}{}^{d} V^{ef} 
\nabla_{a}\nabla^{c}\vp  \nabla^{a}\vp  \nabla^{b}\vp   \nn\\&&+ 
e^{\vp }  
 c_{82}  R_{efbd} V_{c}{}^{d} V^{ef} 
\nabla_{a}\nabla^{c}\vp  \nabla^{a}\vp  \nabla^{b}\vp  + 
e^{\vp }  
 c_{129}  R_{dfbe} V_{a}{}^{e} V_{c}{}^{f} 
\nabla^{a}\vp  \nabla^{b}\vp  \nabla^{c}\nabla^{d}\vp  \nn\\&& + 
e^{\vp }  
 c_{131}  R_{efbd} V_{a}{}^{e} V_{c}{}^{f} 
\nabla^{a}\vp  \nabla^{b}\vp  \nabla^{c}\nabla^{d}\vp  + 
e^{\vp }  
 c_{128}  R_{cfbd} V_{a}{}^{e} V_{e}{}^{f} 
\nabla^{a}\vp  \nabla^{b}\vp  \nabla^{c}\nabla^{d}\vp  \nn\\&& + 
e^{\vp }  
 c_{132}  R_{efbd} V_{ac} V^{ef} \nabla^{a}\vp  
\nabla^{b}\vp  \nabla^{c}\nabla^{d}\vp  \,,
\eeqa
\beqa
[R^2\vp''^2]_5&=&c_{17}  R_{b}{}^{def} R_{efcd} 
\nabla_{a}\nabla^{c}\vp  \nabla^{a}\nabla^{b}\vp  +  
 c_{38}  R^{cdef} R_{efcd} 
\nabla^{a}\nabla^{b}\vp  \nabla_{b}\nabla_{a}\vp   \nn\\&&+  
 c_{121}  R_{a}{}^{e}{}_{b}{}^{f} R_{cfde} 
\nabla^{a}\nabla^{b}\vp  \nabla^{c}\nabla^{d}\vp  +  
 c_{122}  R_{a}{}^{e}{}_{c}{}^{f} R_{dfbe} 
\nabla^{a}\nabla^{b}\vp  \nabla^{c}\nabla^{d}\vp  \nn\\&& +  
 c_{125}  R_{efbd} R^{ef}{}_{ac} 
\nabla^{a}\nabla^{b}\vp  \nabla^{c}\nabla^{d}\vp ,
\eeqa
\beqa
[R^2\vp'^2\vp'']_4&=&c_{84}  R_{b}{}^{def} R_{efcd} 
\nabla_{a}\nabla^{c}\vp  \nabla^{a}\vp  \nabla^{b}\vp  + 
  c_{88}  R^{cdef} R_{efcd} 
\nabla^{a}\vp  \nabla_{b}\nabla_{a}\vp  \nabla^{b}\vp \nn\\&& + 
  c_{130}  R_{c}{}^{e}{}_{a}{}^{f} R_{dfbe} 
\nabla^{a}\vp  \nabla^{b}\vp  \nabla^{c}\nabla^{d}\vp  + 
  c_{133}  R_{efbd} R^{ef}{}_{ac} 
\nabla^{a}\vp  \nabla^{b}\vp  \nabla^{c}\nabla^{d}\vp \,,
\eeqa
\beqa
[R^2\vp'^4]_3&=&c_{87}  R^{cdef} R_{efcd} \nabla_{a}\vp  
\nabla^{a}\vp  \nabla_{b}\vp  \nabla^{b}\vp  +  
 c_{112}  R_{b}{}^{def} R_{efcd} 
\nabla_{a}\vp  \nabla^{a}\vp  \nabla^{b}\vp  
\nabla^{c}\vp \nn\\&& +  
 c_{200}  R_{a}{}^{e}{}_{b}{}^{f} R_{cfde} 
\nabla^{a}\vp  \nabla^{b}\vp  \nabla^{c}\vp  
\nabla^{d}\vp \,,
\eeqa
\beqa
[V^3V'\vp'^3]_{5}&=&e^{2 \vp }  
 c_{97}  V_{c}{}^{e} V_{d}{}^{f} V_{ef} \nabla_{a}\vp  
\nabla^{a}\vp  \nabla^{b}\vp  \nabla^{c}V_{b}{}^{d} + e^{2 
\vp }  
 c_{100}  V_{bd} V_{c}{}^{f} V_{ef} \nabla_{a}\vp  
\nabla^{a}\vp  \nabla^{b}\vp  \nabla^{c}V^{de} \nn\\&&+ e^{2 \vp } 
 c_{
   104}  V_{c}{}^{e} V_{d}{}^{f} V_{ef} \nabla_{a}V_{b}{}^{d} 
\nabla^{a}\vp  \nabla^{b}\vp  \nabla^{c}\vp  + e^{2 \vp } 
 c_{
   106}  V_{bd} V_{c}{}^{f} V_{ef} \nabla_{a}V^{de} \nabla^{a}
\vp  \nabla^{b}\vp  \nabla^{c}\vp \nn\\&& + e^{2 \vp }  
 c_{150}  V_{bd} V_{c}{}^{f} V_{ef} \nabla^{a}\vp  
\nabla^{b}\vp  \nabla^{c}\vp  \nabla^{d}V_{a}{}^{e}\,,
\eeqa
\beqa
[R^2V^4]_{13}&=&e^{2 \vp }  
 c_{170}  R_{d}{}^{fgh} R_{ghef} 
V_{a}{}^{c} V^{ab} V_{b}{}^{d} V_{c}{}^{e} + e^{2 \vp }  
 c_{37}  R_{b}{}^{g}{}_{e}{}^{h} R_{fhcg} 
V_{a}{}^{c} V^{ab} V_{d}{}^{f} V^{de} \nn\\&&+ e^{2 \vp }  
 c_{257}  R_{ghcf} R^{gh}{}_{be} 
V_{a}{}^{c} V^{ab} V_{d}{}^{f} V^{de} + e^{2 \vp }  
 c_{66}  R_{e}{}^{g}{}_{c}{}^{h} R_{fhdg} 
V_{a}{}^{c} V^{ab} V_{b}{}^{d} V^{ef}\nn\\&& + e^{2 \vp }  
 c_{258}  R_{ghef} R^{gh}{}_{cd} 
V_{a}{}^{c} V^{ab} V_{b}{}^{d} V^{ef} + e^{2 \vp }  
 c_{259}  R_{ghdf} R^{gh}{}_{ce} 
V_{a}{}^{c} V^{ab} V_{b}{}^{d} V^{ef} \nn\\&&+ e^{2 \vp }  
 c_{47}  R_{d}{}^{h}{}_{be} R_{fhcg} 
V_{a}{}^{c} V^{ab} V^{de} V^{fg} + e^{2 \vp }  
 c_{75}  R_{b}{}^{h}{}_{cd} R_{fheg} 
V_{a}{}^{c} V^{ab} V^{de} V^{fg} \nn\\&&+ e^{2 \vp }  
 c_{144}  R_{bdf}{}^{h} R_{ghce} 
V_{a}{}^{c} V^{ab} V^{de} V^{fg} + e^{2 \vp }  
 c_{152}  R_{d}{}^{h}{}_{bf} R_{ghce} 
V_{a}{}^{c} V^{ab} V^{de} V^{fg} \nn\\&&+ e^{2 \vp }  
 c_{10}  R_{acbd} R_{egfh} V^{ab} V^{cd} 
V^{ef} V^{gh} + e^{2 \vp }  
 c_{27}  R_{aceg} R_{fhbd} V^{ab} V^{cd} 
V^{ef} V^{gh}\nn\\&& + e^{2 \vp }  
 c_{160}  R_{abce} R_{ghdf} V^{ab} V^{cd} 
V^{ef} V^{gh},
\eeqa
\beqa
[R^2V^2\vp'']_{13}&=&e^{\vp }  
 c_{8}  R^{defg} R_{fgde} V_{a}{}^{c} 
V_{bc} \nabla^{a}\nabla^{b}\vp  + e^{\vp }  
 c_{6}  R_{c}{}^{efg} R_{fgde} V_{a}{}^{c} 
V_{b}{}^{d} \nabla^{a}\nabla^{b}\vp \nn\\&& + e^{\vp }  
 c_{5}  R_{b}{}^{efg} R_{fgde} V_{a}{}^{c} 
V_{c}{}^{d} \nabla^{a}\nabla^{b}\vp  + e^{\vp }  
 c_{292}  R_{a}{}^{f}{}_{b}{}^{g} R_{dgef} 
V_{c}{}^{e} V^{cd} \nabla^{a}\nabla^{b}\vp \nn\\&& + e^{\vp }  
 c_{294}  R_{d}{}^{f}{}_{a}{}^{g} R_{egbf} 
V_{c}{}^{e} V^{cd} \nabla^{a}\nabla^{b}\vp  + e^{\vp }  
 c_{9}  R_{fgbe} R^{fg}{}_{ad} V_{c}{}^{e} 
V^{cd} \nabla^{a}\nabla^{b}\vp \nn\\&& + e^{\vp }  
 c_{295}  R_{d}{}^{f}{}_{b}{}^{g} R_{egcf} 
V_{a}{}^{c} V^{de} \nabla^{a}\nabla^{b}\vp  + e^{\vp }  
 c_{11}  R_{fgde} R^{fg}{}_{bc} V_{a}{}^{c} 
V^{de} \nabla^{a}\nabla^{b}\vp \nn\\&& + e^{\vp }  
 c_{12}  R_{fgce} R^{fg}{}_{bd} V_{a}{}^{c} 
V^{de} \nabla^{a}\nabla^{b}\vp  + e^{\vp }  
 c_{293}  R_{c}{}^{g}{}_{ad} R_{egbf} 
V^{cd} V^{ef} \nabla^{a}\nabla^{b}\vp \nn\\&& + e^{\vp }  
 c_{296}  R_{a}{}^{g}{}_{bc} R_{egdf} 
V^{cd} V^{ef} \nabla^{a}\nabla^{b}\vp  + e^{\vp }  
 c_{1}  R_{ace}{}^{g} R_{fgbd} V^{cd} 
V^{ef} \nabla^{a}\nabla^{b}\vp \nn\\&& + e^{\vp }  
 c_{2}  R_{c}{}^{g}{}_{ae} R_{fgbd} V^{cd} 
V^{ef} \nabla^{a}\nabla^{b}\vp  \,,
\eeqa
\beqa
[R^2V^2\vp'^2]_{17}&=&e^{\vp }  
 c_{285}  R_{c}{}^{efg} R_{fgde} 
V_{b}{}^{d} V^{bc} \nabla_{a}\vp  \nabla^{a}\vp  + e^{\vp }  
 c_{281}  R_{b}{}^{f}{}_{d}{}^{g} 
R_{egcf} V^{bc} V^{de} \nabla_{a}\vp  \nabla^{a}\vp \nn\\&& + 
e^{\vp }  
 c_{286}  R_{fgde} R^{fg}{}_{bc} V^{bc} 
V^{de} \nabla_{a}\vp  \nabla^{a}\vp  + e^{\vp }  
 c_{287}  R_{fgce} R^{fg}{}_{bd} V^{bc} 
V^{de} \nabla_{a}\vp  \nabla^{a}\vp \nn\\&& + e^{\vp }  
 c_{76}  R^{defg} R_{fgde} V_{a}{}^{c} 
V_{bc} \nabla^{a}\vp  \nabla^{b}\vp  + e^{\vp }  
 c_{73}  R_{c}{}^{efg} R_{fgde} V_{a}{}^{c} 
V_{b}{}^{d} \nabla^{a}\vp  \nabla^{b}\vp  \nn\\&&+ e^{\vp }  
 c_{72}  R_{b}{}^{efg} R_{fgde} V_{a}{}^{c} 
V_{c}{}^{d} \nabla^{a}\vp  \nabla^{b}\vp  + e^{\vp }  
 c_{60}  R_{a}{}^{f}{}_{b}{}^{g} R_{dgef} 
V_{c}{}^{e} V^{cd} \nabla^{a}\vp  \nabla^{b}\vp \nn\\&& + e^{\vp }  
 c_{62}  R_{d}{}^{f}{}_{a}{}^{g} R_{egbf} 
V_{c}{}^{e} V^{cd} \nabla^{a}\vp  \nabla^{b}\vp  + e^{\vp }  
 c_{77}  R_{fgbe} R^{fg}{}_{ad} 
V_{c}{}^{e} V^{cd} \nabla^{a}\vp  \nabla^{b}\vp  \nn\\&&+ e^{\vp }  
 c_{63}  R_{d}{}^{f}{}_{b}{}^{g} R_{egcf} 
V_{a}{}^{c} V^{de} \nabla^{a}\vp  \nabla^{b}\vp  + e^{\vp }  
 c_{78}  R_{fgde} R^{fg}{}_{bc} 
V_{a}{}^{c} V^{de} \nabla^{a}\vp  \nabla^{b}\vp \nn\\&& + e^{\vp }  
 c_{79}  R_{fgce} R^{fg}{}_{bd} 
V_{a}{}^{c} V^{de} \nabla^{a}\vp  \nabla^{b}\vp  + e^{\vp }  
 c_{61}  R_{c}{}^{g}{}_{ad} R_{egbf} 
V^{cd} V^{ef} \nabla^{a}\vp  \nabla^{b}\vp  \nn\\&&+ e^{\vp }  
 c_{64}  R_{a}{}^{g}{}_{bc} R_{egdf} V^{cd} 
V^{ef} \nabla^{a}\vp  \nabla^{b}\vp  + e^{\vp }  
 c_{68}  R_{ace}{}^{g} R_{fgbd} V^{cd} 
V^{ef} \nabla^{a}\vp  \nabla^{b}\vp \nn\\&& + e^{\vp }  
 c_{69}  R_{c}{}^{g}{}_{ae} R_{fgbd} V^{cd} 
V^{ef} \nabla^{a}\vp  \nabla^{b}\vp ,
\eeqa
\beqa
[RV^4\vp'^2]_{9}&=&e^{2 \vp }  
 c_{280}  R_{cfdg} V_{b}{}^{d} V^{bc} V_{e}{}^{g} 
V^{ef} \nabla_{a}\vp  \nabla^{a}\vp  + e^{2 \vp }  
 c_{284}  R_{fgde} V_{b}{}^{d} V^{bc} V_{c}{}^{e} 
V^{fg} \nabla_{a}\vp  \nabla^{a}\vp   \nn\\&&+ e^{2 \vp }  
 c_{56}  R_{afbg} V_{c}{}^{e} V^{cd} V_{d}{}^{f} 
V_{e}{}^{g} \nabla^{a}\vp  \nabla^{b}\vp  + e^{2 \vp }  
 c_{67}  R_{fgbc} V_{a}{}^{c} V_{d}{}^{f} V^{de} 
V_{e}{}^{g} \nabla^{a}\vp  \nabla^{b}\vp  \nn\\&& + e^{2 \vp }  
 c_{57}  R_{cfdg} V_{a}{}^{c} V_{b}{}^{d} 
V_{e}{}^{g} V^{ef} \nabla^{a}\vp  \nabla^{b}\vp  + e^{2 \vp } 
 c_{
   58}  R_{dfbg} V_{a}{}^{c} V_{c}{}^{d} V_{e}{}^{g} 
V^{ef} \nabla^{a}\vp  \nabla^{b}\vp  \nn\\&& + e^{2 \vp }  
 c_{71}  R_{fgde} V_{a}{}^{c} V_{b}{}^{d} 
V_{c}{}^{e} V^{fg} \nabla^{a}\vp  \nabla^{b}\vp  + e^{2 \vp } 
 c_{
   70}  R_{fgbe} V_{a}{}^{c} V_{c}{}^{d} V_{d}{}^{e} 
V^{fg} \nabla^{a}\vp  \nabla^{b}\vp  \nn\\&& + e^{2 \vp }  
 c_{59}  R_{dfeg} V_{a}{}^{c} V_{bc} V^{de} V^{fg} 
\nabla^{a}\vp  \nabla^{b}\vp 
\,,
\eeqa
\beqa
[RV^2V'^2]_{27}&=&e^{2 \vp }  
 c_{270}  R_{fgde} V_{c}{}^{e} V^{fg} 
\nabla_{a}V_{b}{}^{d} \nabla^{a}V^{bc} + e^{2 \vp }  
 c_{276}  R_{fgde} V_{b}{}^{f} V_{c}{}^{g} 
\nabla_{a}V^{de} \nabla^{a}V^{bc}  \nn\\&&+ e^{2 \vp }  
 c_{274}  R_{fgce} V_{b}{}^{f} V_{d}{}^{g} 
\nabla_{a}V^{de} \nabla^{a}V^{bc} + e^{2 \vp }  
 c_{275}  R_{fgce} V_{bd} V^{fg} \nabla_{a}V^{de} 
\nabla^{a}V^{bc} \nn\\&& + e^{2 \vp }  
 c_{19}  R_{dfeg} V^{de} V^{fg} \nabla^{a}V^{bc} 
\nabla_{b}V_{ac} + e^{2 \vp }  
 c_{22}  R_{cfdg} V_{e}{}^{g} V^{ef} 
\nabla^{a}V^{bc} \nabla_{b}V_{a}{}^{d}  \nn\\&&+ e^{2 \vp }  
 c_{23}  R_{fgde} V_{c}{}^{e} V^{fg} 
\nabla^{a}V^{bc} \nabla_{b}V_{a}{}^{d} + e^{2 \vp }  
 c_{25}  R_{dfag} V_{e}{}^{g} V^{ef} 
\nabla^{a}V^{bc} \nabla_{b}V_{c}{}^{d}  \nn\\&&+ e^{2 \vp }  
 c_{30}  R_{egcf} V_{a}{}^{f} V_{d}{}^{g} 
\nabla^{a}V^{bc} \nabla_{b}V^{de} + e^{2 \vp }  
 c_{32}  R_{fgce} V_{a}{}^{f} V_{d}{}^{g} 
\nabla^{a}V^{bc} \nabla_{b}V^{de} \nn\\&& + e^{2 \vp }  
 c_{29}  R_{egaf} V_{c}{}^{f} V_{d}{}^{g} 
\nabla^{a}V^{bc} \nabla_{b}V^{de} + e^{2 \vp }  
 c_{28}  R_{egac} V_{d}{}^{f} V_{f}{}^{g} 
\nabla^{a}V^{bc} \nabla_{b}V^{de}  \nn\\&&+ e^{2 \vp }  
 c_{33}  R_{fgce} V_{ad} V^{fg} \nabla^{a}V^{bc} 
\nabla_{b}V^{de} + e^{2 \vp }  
 c_{165}  R_{egdf} V_{a}{}^{f} V_{c}{}^{g} 
\nabla^{a}V^{bc} \nabla^{d}V_{b}{}^{e}  \nn\\&&+ e^{2 \vp }  
 c_{167}  R_{fgce} V_{a}{}^{f} V_{d}{}^{g} 
\nabla^{a}V^{bc} \nabla^{d}V_{b}{}^{e} + e^{2 \vp }  
 c_{161}  R_{dgcf} V_{a}{}^{f} V_{e}{}^{g} 
\nabla^{a}V^{bc} \nabla^{d}V_{b}{}^{e} \nn\\&& + e^{2 \vp }  
 c_{159}  R_{dgce} V_{a}{}^{f} V_{f}{}^{g} 
\nabla^{a}V^{bc} \nabla^{d}V_{b}{}^{e} + e^{2 \vp }  
 c_{163}  R_{egcd} V_{a}{}^{f} V_{f}{}^{g} 
\nabla^{a}V^{bc} \nabla^{d}V_{b}{}^{e}  \nn\\&&+ e^{2 \vp }  
 c_{162}  R_{egad} V_{c}{}^{f} V_{f}{}^{g} 
\nabla^{a}V^{bc} \nabla^{d}V_{b}{}^{e} + e^{2 \vp }  
 c_{168}  R_{fgce} V_{ad} V^{fg} \nabla^{a}V^{bc} 
\nabla^{d}V_{b}{}^{e}  \nn\\&&+ e^{2 \vp }  
 c_{189}  R_{fgcd} V_{a}{}^{g} V_{be} 
\nabla^{a}V^{bc} \nabla^{d}V^{ef} + e^{2 \vp }  
 c_{184}  R_{efcd} V_{a}{}^{g} V_{bg} 
\nabla^{a}V^{bc} \nabla^{d}V^{ef}  \nn\\&&+ e^{2 \vp }  
 c_{185}  R_{egcf} V_{ad} V_{b}{}^{g} 
\nabla^{a}V^{bc} \nabla^{d}V^{ef} + e^{2 \vp }  
 c_{190}  R_{fgcd} V_{ae} V_{b}{}^{g} 
\nabla^{a}V^{bc} \nabla^{d}V^{ef}  \nn\\&&+ e^{2 \vp }  
 c_{187}  R_{fgbc} V_{ae} V_{d}{}^{g} 
\nabla^{a}V^{bc} \nabla^{d}V^{ef} + e^{2 \vp }  
 c_{183}  R_{dfbc} V_{a}{}^{g} V_{eg} 
\nabla^{a}V^{bc} \nabla^{d}V^{ef} \nn\\&& + e^{2 \vp }  
 c_{182}  R_{cfad} V_{b}{}^{g} V_{eg} 
\nabla^{a}V^{bc} \nabla^{d}V^{ef}\,,
\eeqa
\beqa
[RV^6]_1&=&e^{3 \vp }  
 c_{127}  R_{ghef} V_{a}{}^{c} V^{ab} V_{b}{}^{d} 
V_{c}{}^{e} V_{d}{}^{f} V^{gh},
\eeqa
\beqa
[V^4\vp'^2\vp'']_1&=&e^{2 \vp }  
 c_{126}  V_{ac} V_{b}{}^{e} V_{d}{}^{f} V_{ef} 
\nabla^{a}\vp  \nabla^{b}\vp  \nabla^{c}\nabla^{d}\vp 
\,,
\eeqa
\beqa
[R^2VV'\vp']_{14}&=&e^{\vp }  
 c_{279}  R_{c}{}^{efg} R_{fgde} 
V_{b}{}^{d} \nabla_{a}V^{bc} \nabla^{a}\vp  + e^{\vp }  
 c_{41}  R_{c}{}^{efg} R_{fgde} V_{b}{}^{d} 
\nabla^{a}\vp  \nabla^{b}V_{a}{}^{c} \nn\\&&+ e^{\vp }  
 c_{42}  R_{fgde} R^{fg}{}_{bc} V^{de} 
\nabla^{a}\vp  \nabla^{b}V_{a}{}^{c} + e^{\vp }  
 c_{43}  R_{fgce} R^{fg}{}_{bd} V^{de} 
\nabla^{a}\vp  \nabla^{b}V_{a}{}^{c}\nn\\&& + e^{\vp }  
 c_{51}  R_{b}{}^{efg} R_{fgde} V_{ac} 
\nabla^{a}\vp  \nabla^{b}V^{cd} + e^{\vp }  
 c_{45}  R_{c}{}^{f}{}_{b}{}^{g} R_{egdf} 
V_{a}{}^{e} \nabla^{a}\vp  \nabla^{b}V^{cd} \nn\\&&+ e^{\vp }  
 c_{54}  R_{fgcd} R^{fg}{}_{be} V_{a}{}^{e} 
\nabla^{a}\vp  \nabla^{b}V^{cd} + e^{\vp }  
 c_{53}  R_{fgcd} R^{fg}{}_{ae} V_{b}{}^{e} 
\nabla^{a}\vp  \nabla^{b}V^{cd}\nn\\&& + e^{\vp }  
 c_{52}  R_{fgbe} R^{fg}{}_{ad} V_{c}{}^{e} 
\nabla^{a}\vp  \nabla^{b}V^{cd} + e^{\vp }  
 c_{44}  R_{b}{}^{g}{}_{ac} R_{egdf} V^{ef} 
\nabla^{a}\vp  \nabla^{b}V^{cd} \nn\\&&+ e^{\vp }  
 c_{46}  R_{c}{}^{g}{}_{ab} R_{egdf} V^{ef} 
\nabla^{a}\vp  \nabla^{b}V^{cd} + e^{\vp }  
 c_{48}  R_{c}{}^{g}{}_{be} R_{fgad} V^{ef} 
\nabla^{a}\vp  \nabla^{b}V^{cd} \nn\\&&+ e^{\vp }  
 c_{49}  R_{b}{}^{g}{}_{ae} R_{fgcd} V^{ef} 
\nabla^{a}\vp  \nabla^{b}V^{cd} + e^{\vp }  
 c_{50}  R_{e}{}^{g}{}_{ab} R_{fgcd} V^{ef} 
\nabla^{a}\vp  \nabla^{b}V^{cd},
\eeqa
\beqa
[V^4\vp'^4]_{2}&=&e^{2 \vp }  
 c_{108}  V_{b}{}^{d} V_{c}{}^{e} V_{d}{}^{f} V_{ef} 
\nabla_{a}\vp  \nabla^{a}\vp  \nabla^{b}\vp  
\nabla^{c}\vp \nn\\&& + e^{2 \vp }  
 c_{198}  V_{a}{}^{e} V_{be} V_{c}{}^{f} V_{df} 
\nabla^{a}\vp  \nabla^{b}\vp  \nabla^{c}\vp  
\nabla^{d}\vp ,
\eeqa
\beqa
[RV^2\vp'^4]_4&=&e^{\vp }  
 c_{86}  R_{cedf} V^{cd} V^{ef} \nabla_{a}\vp  
\nabla^{a}\vp  \nabla_{b}\vp  \nabla^{b}\vp  + e^{\vp } 
  c_{110}  R_{becf} V_{d}{}^{f} V^{de} 
\nabla_{a}\vp  \nabla^{a}\vp  \nabla^{b}\vp  
\nabla^{c}\vp  \nn\\&& + e^{\vp }  
 c_{111}  R_{efcd} V_{b}{}^{d} V^{ef} 
\nabla_{a}\vp  \nabla^{a}\vp  \nabla^{b}\vp  
\nabla^{c}\vp  \nn\\&&+ e^{\vp }  
 c_{199}  R_{cedf} V_{a}{}^{e} V_{b}{}^{f} 
\nabla^{a}\vp  \nabla^{b}\vp  \nabla^{c}\vp  
\nabla^{d}\vp \,,
\eeqa
\beqa
[RVV'\vp'^3]_{10}&=&e^{\vp }  
 c_{98}  R_{efcd} V^{ef} \nabla_{a}\vp  
\nabla^{a}\vp  \nabla^{b}\vp  \nabla^{c}V_{b}{}^{d} + e^{\vp 
} 
  c_{102}  R_{dfce} V_{b}{}^{f} \nabla_{a}\vp  
\nabla^{a}\vp  \nabla^{b}\vp  \nabla^{c}V^{de} \nn\\&&+ e^{\vp } 
  c_{101}  R_{dfbe} V_{c}{}^{f} \nabla_{a}\vp  
\nabla^{a}\vp  \nabla^{b}\vp  \nabla^{c}V^{de} + e^{\vp } 
  c_{103}  R_{efbc} V_{d}{}^{f} \nabla_{a}\vp  
\nabla^{a}\vp  \nabla^{b}\vp  \nabla^{c}V^{de} \nn\\&&+ e^{\vp } 
  c_{105}  R_{efcd} V^{ef} \nabla_{a}V_{b}{}^{d} 
\nabla^{a}\vp  \nabla^{b}\vp  \nabla^{c}\vp  + e^{\vp } 
  c_{107}  R_{bfce} V_{d}{}^{f} \nabla_{a}V^{de} 
\nabla^{a}\vp  \nabla^{b}\vp  \nabla^{c}\vp \nn\\&& + e^{\vp } 
  c_{153}  R_{dfce} V_{b}{}^{f} \nabla^{a}\vp  
\nabla^{b}\vp  \nabla^{c}\vp  \nabla^{d}V_{a}{}^{e} + e^{\vp 
} 
  c_{154}  R_{efcd} V_{b}{}^{f} \nabla^{a}\vp  
\nabla^{b}\vp  \nabla^{c}\vp  \nabla^{d}V_{a}{}^{e} \nn\\&&+ e^{\vp 
} 
  c_{151}  R_{bfce} V_{d}{}^{f} \nabla^{a}\vp  
\nabla^{b}\vp  \nabla^{c}\vp  \nabla^{d}V_{a}{}^{e} \nn\\&&+ e^{\vp 
} 
  c_{195}  R_{bfcd} V_{ae} \nabla^{a}\vp  \nabla^{b}
\vp  \nabla^{c}\vp  \nabla^{d}V^{ef}, 
\eeqa
\beqa
[RV^4\vp'']_{7}&=&e^{2 \vp }  
 c_{288}  R_{afbg} V_{c}{}^{e} V^{cd} V_{d}{}^{f} 
V_{e}{}^{g} \nabla^{a}\nabla^{b}\vp  + e^{2 \vp }  
 c_{298}  R_{fgbc} V_{a}{}^{c} V_{d}{}^{f} V^{de} 
V_{e}{}^{g} \nabla^{a}\nabla^{b}\vp   \nn\\&&+ e^{2 \vp }  
 c_{289}  R_{cfdg} V_{a}{}^{c} V_{b}{}^{d} 
V_{e}{}^{g} V^{ef} \nabla^{a}\nabla^{b}\vp  + e^{2 \vp }  
 c_{290}  R_{dfbg} V_{a}{}^{c} V_{c}{}^{d} 
V_{e}{}^{g} V^{ef} \nabla^{a}\nabla^{b}\vp  \nn\\&& + e^{2 \vp }  
 c_{4}  R_{fgde} V_{a}{}^{c} V_{b}{}^{d} V_{c}{}^{e} 
V^{fg} \nabla^{a}\nabla^{b}\vp  + e^{2 \vp }  
 c_{3}  R_{fgbe} V_{a}{}^{c} V_{c}{}^{d} V_{d}{}^{e} 
V^{fg} \nabla^{a}\nabla^{b}\vp   \nn\\&&+ e^{2 \vp }  
 c_{291}  R_{dfeg} V_{a}{}^{c} V_{bc} V^{de} V^{fg} 
\nabla^{a}\nabla^{b}\vp ,
\eeqa
\beqa
[V^2\vp'^6]_1&=&e^{\vp }  
 c_{203}  V_{c}{}^{e} V_{de} \nabla_{a}\vp  
\nabla^{a}\vp  \nabla_{b}\vp  \nabla^{b}\vp  
\nabla^{c}\vp  \nabla^{d}\vp ,
\eeqa
\beqa
[V^4\vp''^2]_5&=&e^{2 \vp }  
 c_{14}  V_{b}{}^{d} V_{c}{}^{e} V_{d}{}^{f} V_{ef} 
\nabla_{a}\nabla^{c}\vp  \nabla^{a}\nabla^{b}\vp  + e^{2 
\vp }  
 c_{35}  V_{c}{}^{e} V^{cd} V_{d}{}^{f} V_{ef} 
\nabla^{a}\nabla^{b}\vp  \nabla_{b}\nabla_{a}\vp \nn\\&& + e^{2 
\vp }  
 c_{114}  V_{a}{}^{e} V_{b}{}^{f} V_{ce} V_{df} 
\nabla^{a}\nabla^{b}\vp  \nabla^{c}\nabla^{d}\vp  + e^{2 
\vp }  
 c_{115}  V_{a}{}^{e} V_{be} V_{c}{}^{f} V_{df} 
\nabla^{a}\nabla^{b}\vp  \nabla^{c}\nabla^{d}\vp \nn\\&& + e^{2 
\vp }  
 c_{116}  V_{ac} V_{b}{}^{e} V_{d}{}^{f} V_{ef} 
\nabla^{a}\nabla^{b}\vp  \nabla^{c}\nabla^{d}\vp 
,
\eeqa
\beqa
[RV^2\vp''^2]_8&\!\!\!\!=\!\!\!\!&e^{\vp }  
 c_{15}  R_{becf} V_{d}{}^{f} V^{de} 
\nabla_{a}\nabla^{c}\vp  \nabla^{a}\nabla^{b}\vp  + e^{\vp 
} 
  c_{16}  R_{efcd} V_{b}{}^{d} V^{ef} 
\nabla_{a}\nabla^{c}\vp  \nabla^{a}\nabla^{b}\vp  \nn\\&&+ e^{\vp 
} 
  c_{36}  R_{cedf} V^{cd} V^{ef} 
\nabla^{a}\nabla^{b}\vp  \nabla_{b}\nabla_{a}\vp  + e^{\vp 
} 
  c_{119}  R_{cedf} V_{a}{}^{e} V_{b}{}^{f} 
\nabla^{a}\nabla^{b}\vp  \nabla^{c}\nabla^{d}\vp \nn\\&& + e^{\vp 
} 
  c_{118}  R_{bfde} V_{a}{}^{e} V_{c}{}^{f} 
\nabla^{a}\nabla^{b}\vp  \nabla^{c}\nabla^{d}\vp  + e^{\vp 
} 
  c_{123}  R_{efbd} V_{a}{}^{e} V_{c}{}^{f} 
\nabla^{a}\nabla^{b}\vp  \nabla^{c}\nabla^{d}\vp  \nn\\&&+ e^{\vp 
} 
  c_{120}  R_{cfbd} V_{a}{}^{e} V_{e}{}^{f} 
\nabla^{a}\nabla^{b}\vp  \nabla^{c}\nabla^{d}\vp  + e^{\vp 
} 
  c_{124}  R_{efbd} V_{ac} V^{ef} \nabla^{a}\nabla^{b}
\vp  \nabla^{c}\nabla^{d}\vp  ,
\eeqa
\beqa
[V^2V'^2\vp'']_{15}&=&e^{2 \vp }  
 c_{34}  V_{cd} V_{ef} \nabla_{a}\nabla^{f}\vp  
\nabla^{a}V^{bc} \nabla_{b}V^{de} + e^{2 \vp }  
 c_{39}  V_{cd} V_{ef} \nabla_{a}V^{de} \nabla^{a}V^{bc} 
\nabla_{b}\nabla^{f}\vp \nn\\&& + e^{2 \vp }  
 c_{92}  V_{d}{}^{f} V_{ef} \nabla_{a}V_{b}{}^{d} 
\nabla^{a}V^{bc} \nabla_{c}\nabla^{e}\vp  + e^{2 \vp }  
 c_{93}  V_{d}{}^{f} V_{ef} \nabla^{a}V^{bc} 
\nabla_{b}V_{a}{}^{d} \nabla_{c}\nabla^{e}\vp \nn\\&& + e^{2 \vp } 
  c_{141}  V_{c}{}^{f} V_{ef} \nabla^{a}V^{bc} 
\nabla_{b}V^{de} \nabla_{d}\nabla_{a}\vp  + e^{2 \vp }  
 c_{142}  V_{c}{}^{f} V_{ef} \nabla_{a}V^{de} 
\nabla^{a}V^{bc} \nabla_{d}\nabla_{b}\vp \nn\\&& + e^{2 \vp }  
 c_{171}  V_{cd} V_{ef} \nabla_{a}\nabla^{f}\vp  
\nabla^{a}V^{bc} \nabla^{d}V_{b}{}^{e} + e^{2 \vp }  
 c_{174}  V_{d}{}^{f} V_{ef} \nabla^{a}V^{bc} 
\nabla_{c}\nabla_{a}\vp  \nabla^{d}V_{b}{}^{e}\nn\\&& + e^{2 \vp } 
  c_{175}  V_{ad} V_{ef} \nabla^{a}V^{bc} 
\nabla_{c}\nabla^{f}\vp  \nabla^{d}V_{b}{}^{e} + e^{2 \vp } 
  c_{178}  V_{c}{}^{f} V_{ef} \nabla^{a}V^{bc} 
\nabla_{d}\nabla_{a}\vp  \nabla^{d}V_{b}{}^{e} \nn\\&&+ e^{2 \vp } 
  c_{197}  V_{be} V_{cf} \nabla^{a}V^{bc} 
\nabla_{d}\nabla_{a}\vp  \nabla^{d}V^{ef} + e^{2 \vp }  
 c_{209}  V_{d}{}^{f} V_{ef} \nabla^{a}V^{bc} 
\nabla_{b}V_{ac} \nabla^{d}\nabla^{e}\vp  \nn\\&&+ e^{2 \vp }  
 c_{219}  V_{ad} V_{cf} \nabla^{a}V^{bc} \nabla^{d}V^{ef} 
\nabla_{e}\nabla_{b}\vp  + e^{2 \vp }  
 c_{245}  V_{ce} V_{df} \nabla_{a}V_{b}{}^{d} 
\nabla^{a}V^{bc} \nabla^{e}\nabla^{f}\vp \nn\\&& + e^{2 \vp }  
 c_{246}  V_{ce} V_{df} \nabla^{a}V^{bc} 
\nabla_{b}V_{a}{}^{d} \nabla^{e}\nabla^{f}\vp ,
\eeqa
\beqa
[RV^3V'\vp']_{1}&=&e^{2 \vp }  
 c_{40}  R_{fgde} V_{b}{}^{d} V_{c}{}^{e} V^{fg} 
\nabla^{a}\vp  \nabla^{b}V_{a}{}^{c},
\eeqa
\beqa
[RVV'\vp'\vp'']_2&\!\!\!\!=\!\!\!\!&e^{\vp }  
 c_{55}  R_{efcd} V_{b}{}^{f} \nabla_{a}\nabla^{e}
\vp  \nabla^{a}\vp  \nabla^{b}V^{cd} + e^{\vp }  
 c_{210}  R_{efbc} V_{d}{}^{f} \nabla^{a}\vp  
\nabla^{b}V_{a}{}^{c} \nabla^{d}\nabla^{e}\vp ,
\eeqa
\beqa
[V^2\vp'^2\vp''^2]_2&=&e^{\vp }  
 c_{134}  V_{c}{}^{e} V_{de} \nabla^{a}\vp  
\nabla_{b}\nabla_{a}\vp  \nabla^{b}\vp  \nabla^{c}\nabla^{d}
\vp \nn\\&& + e^{\vp }  
 c_{211}  V_{bd} V_{ce} \nabla_{a}\nabla^{c}\vp  
\nabla^{a}\vp  \nabla^{b}\vp  \nabla^{d}\nabla^{e}\vp ,
\eeqa
\beqa
[V^2V'^2\vp'^2]_{11}&=&e^{2 \vp }  
 c_{89}  V_{d}{}^{f} V_{ef} \nabla_{a}\vp  
\nabla^{a}\vp  \nabla^{b}V^{cd} \nabla_{c}V_{b}{}^{e} + e^{2 
\vp }  
 c_{90}  V_{d}{}^{f} V_{ef} \nabla_{a}V^{cd} 
\nabla^{a}\vp  \nabla^{b}\vp  \nabla_{c}V_{b}{}^{e}  \nn\\&&+ e^{2 
\vp }  
 c_{95}  V_{d}{}^{f} V_{ef} \nabla^{a}\vp  
\nabla^{b}\vp  \nabla_{c}V_{b}{}^{e} \nabla^{c}V_{a}{}^{d} + 
e^{2 \vp }  
 c_{96}  V_{be} V_{df} \nabla^{a}\vp  \nabla^{b}\vp  
\nabla_{c}V^{ef} \nabla^{c}V_{a}{}^{d}  \nn\\&&+ e^{2 \vp }  
 c_{137}  V_{b}{}^{f} V_{ef} \nabla^{a}\vp  
\nabla^{b}\vp  \nabla^{c}V_{a}{}^{d} \nabla_{d}V_{c}{}^{e} + 
e^{2 \vp }  
 c_{179}  V_{d}{}^{f} V_{ef} \nabla_{a}V_{b}{}^{c} 
\nabla^{a}\vp  \nabla^{b}\vp  \nabla^{d}V_{c}{}^{e}  \nn\\&&+ e^{2 
\vp }  
 c_{230}  V_{d}{}^{f} V_{ef} \nabla_{a}V^{cd} 
\nabla^{a}\vp  \nabla^{b}\vp  \nabla^{e}V_{bc} + e^{2 \vp } 
 c_{
   232}  V_{ce} V_{df} \nabla_{a}V^{cd} \nabla^{a}\vp  
\nabla^{b}\vp  \nabla^{e}V_{b}{}^{f}  \nn\\&&+ e^{2 \vp }  
 c_{235}  V_{ce} V_{df} \nabla^{a}\vp  \nabla^{b}\vp  
\nabla^{c}V_{a}{}^{d} \nabla^{e}V_{b}{}^{f} + e^{2 \vp }  
 c_{237}  V_{b}{}^{f} V_{ef} \nabla^{a}\vp  
\nabla^{b}\vp  \nabla^{c}V_{a}{}^{d} \nabla^{e}V_{cd} \nn\\&& + e^{2 
\vp }  
 c_{241}  V_{be} V_{df} \nabla^{a}\vp  \nabla^{b}\vp  
\nabla^{c}V_{a}{}^{d} \nabla^{e}V_{c}{}^{f},
\eeqa
\beqa
[VV'\vp'\vp''^2]_{1}&=&e^{\vp }  
 c_{94}  V_{de} \nabla^{a}\vp  \nabla_{b}\nabla_{a}\vp 
 \nabla^{b}V^{cd} \nabla_{c}\nabla^{e}\vp ,
\eeqa
\beqa
[VV'\vp'^3\vp'']_4&=&e^{\vp }  
 c_{113}  V_{de} \nabla_{a}V_{b}{}^{d} \nabla^{a}\vp  
\nabla^{b}\vp  \nabla_{c}\nabla^{e}\vp  \nabla^{c}\vp  + 
e^{\vp }  
 c_{143}  V_{ce} \nabla_{a}V^{de} \nabla^{a}\vp  
\nabla^{b}\vp  \nabla^{c}\vp  \nabla_{d}\nabla_{b}\vp \nn\\&& + 
e^{\vp }  
 c_{148}  V_{ce} \nabla_{a}V_{b}{}^{d} \nabla^{a}\vp  
\nabla^{b}\vp  \nabla^{c}\vp  \nabla_{d}\nabla^{e}\vp  \nn\\&&+ 
e^{\vp }  
 c_{155}  V_{ce} \nabla^{a}\vp  \nabla^{b}\vp  
\nabla^{c}\vp  \nabla_{d}\nabla_{b}\vp  \nabla^{d}V_{a}{}^{e},
\eeqa
\beqa
[R\vp'^2\vp''^2]_2&=&c_{135}  R_{adbe} \nabla^{a}\vp  \nabla^{b}\vp 
 \nabla_{c}\nabla^{e}\vp  \nabla^{c}\nabla^{d}\vp  +  
 c_{212}  R_{cdbe} \nabla_{a}\nabla^{c}\vp  
\nabla^{a}\vp  \nabla^{b}\vp  \nabla^{d}\nabla^{e}\vp ,
\eeqa
\beqa
[V'^2\vp'^4]_4&=&e^{\vp }  
 c_{136}  \nabla_{a}\vp  \nabla^{a}\vp  
\nabla_{b}\vp  \nabla^{b}\vp  \nabla^{c}V^{de} 
\nabla_{d}V_{ce} + e^{\vp }  
 c_{177}  \nabla_{a}\vp  \nabla^{a}\vp  
\nabla^{b}\vp  \nabla^{c}\vp  \nabla_{d}V_{ce} 
\nabla^{d}V_{b}{}^{e} \nn\\&&+ e^{\vp }  
 c_{205}  \nabla_{a}V_{b}{}^{e} \nabla^{a}\vp  
\nabla^{b}\vp  \nabla_{c}V_{de} \nabla^{c}\vp  
\nabla^{d}\vp  \nn\\&&+ e^{\vp }  
 c_{217}  \nabla_{a}\vp  \nabla^{a}\vp  
\nabla^{b}\vp  \nabla^{c}\vp  \nabla^{d}V_{b}{}^{e} 
\nabla_{e}V_{cd} ,
\eeqa
\beqa
[VV'\vp'^5]_1&=&e^{\vp }  
 c_{201}  V_{de} \nabla_{a}\vp  \nabla^{a}\vp  
\nabla_{b}V_{c}{}^{e} \nabla^{b}\vp  \nabla^{c}\vp  
\nabla^{d}\vp ,
\eeqa
\beqa
[V^2\vp'^4\vp'']_2&=&e^{\vp }  
 c_{204}  V_{c}{}^{e} V_{de} \nabla^{a}\vp  
\nabla_{b}\nabla_{a}\vp  \nabla^{b}\vp  \nabla^{c}\vp  
\nabla^{d}\vp  \nn\\&&+ e^{\vp }  
 c_{214}  V_{bd} V_{ce} \nabla_{a}\vp  \nabla^{a}\vp  
\nabla^{b}\vp  \nabla^{c}\vp  \nabla^{d}\nabla^{e}\vp,
\eeqa
\beqa
[R\vp''^3]_1&=&c_{208}  R_{bdce} \nabla_{a}\nabla^{c}\vp  
\nabla^{a}\nabla^{b}\vp  \nabla^{d}\nabla^{e}\vp , 
\eeqa
\beqa
[R\vp'^4\vp'']_1&=&c_{215}  R_{bdce} \nabla_{a}\vp  \nabla^{a}\vp 
 \nabla^{b}\vp  \nabla^{c}\vp  \nabla^{d}\nabla^{e}\vp,
\eeqa
\beqa
[V'^2\vp''^2]_6&\!\!\!=\!\!\!&e^{\vp }  
 c_{221}  \nabla_{a}V^{de} \nabla^{a}V^{bc} 
\nabla_{d}\nabla_{b}\vp  \nabla_{e}\nabla_{c}\vp  + e^{\vp 
} 
  c_{222}  \nabla^{a}V^{bc} \nabla_{d}\nabla_{a}\vp  
\nabla^{d}V_{b}{}^{e} \nabla_{e}\nabla_{c}\vp \nn\\&& + e^{\vp }  
 c_{224}  \nabla_{a}V_{b}{}^{d} \nabla^{a}V^{bc} 
\nabla_{c}\nabla^{e}\vp  \nabla_{e}\nabla_{d}\vp  + e^{\vp 
} 
  c_{225}  \nabla^{a}V^{bc} \nabla_{b}V_{a}{}^{d} 
\nabla_{c}\nabla^{e}\vp  \nabla_{e}\nabla_{d}\vp \nn\\&& + e^{\vp 
} 
  c_{227}  \nabla^{a}V^{bc} \nabla_{c}\nabla_{a}\vp  
\nabla^{d}V_{b}{}^{e} \nabla_{e}\nabla_{d}\vp  + e^{\vp }  
 c_{229}  \nabla^{a}V^{bc} \nabla_{b}V_{ac} 
\nabla^{d}\nabla^{e}\vp  \nabla_{e}\nabla_{d}\vp ,
\eeqa
\beqa
[V'^2\vp'^2\vp'']_4&=&e^{\vp }  
 c_{223}  \nabla_{a}V^{cd} \nabla^{a}\vp  
\nabla^{b}\vp  \nabla_{c}V_{b}{}^{e} \nabla_{e}\nabla_{d}\vp  
+ e^{\vp }  
 c_{226}  \nabla^{a}\vp  \nabla^{b}\vp  
\nabla_{c}V_{b}{}^{e} \nabla^{c}V_{a}{}^{d} 
\nabla_{e}\nabla_{d}\vp \nn\\&& + e^{\vp }  
 c_{228}  \nabla_{a}V_{b}{}^{c} \nabla^{a}\vp  
\nabla^{b}\vp  \nabla^{d}V_{c}{}^{e} \nabla_{e}\nabla_{d}\vp  
\nn\\&&+ e^{\vp }  
 c_{231}  \nabla_{a}V^{cd} \nabla^{a}\vp  
\nabla^{b}\vp  \nabla_{e}\nabla_{d}\vp  \nabla^{e}V_{bc},
\eeqa
\beqa
[VV'^3\vp']_1&=&e^{2 \vp }  
 c_{251}  V_{ef} \nabla^{a}\vp  \nabla^{b}V_{a}{}^{c} 
\nabla^{d}V_{b}{}^{e} \nabla^{f}V_{cd},\labell{CCPP}
\eeqa
\beqa
[\vp'^4\vp''^2]_2&=&c_{146}  \nabla_{a}\vp  \nabla^{a}\vp  
\nabla_{b}\nabla^{d}\vp  \nabla^{b}\vp  \nabla^{c}\vp  
\nabla_{d}\nabla_{c}\vp \nn\\&& +  
 c_{207}  \nabla^{a}\vp  \nabla_{b}\nabla_{a}\vp  
\nabla^{b}\vp  \nabla^{c}\vp  \nabla_{d}\nabla_{c}\vp  
\nabla^{d}\vp ,
\eeqa
\beqa
[\vp'^8]_1&=&c_{206}  \nabla_{a}\vp  \nabla^{a}\vp  \nabla_{b}\vp 
 \nabla^{b}\vp  \nabla_{c}\vp  \nabla^{c}\vp  
\nabla_{d}\vp  \nabla^{d}\vp ,
\eeqa
\beqa
[V^4V'^2]_{11}&=&e^{3 \vp }  
 c_{269}  V_{c}{}^{e} V_{d}{}^{f} V_{e}{}^{g} V_{fg} 
\nabla_{a}V_{b}{}^{d} \nabla^{a}V^{bc} + e^{3 \vp }  
 c_{272}  V_{b}{}^{f} V_{c}{}^{g} V_{df} V_{eg} 
\nabla_{a}V^{de} \nabla^{a}V^{bc} \nn\\&&+ e^{3 \vp }  
 c_{273}  V_{bd} V_{c}{}^{f} V_{e}{}^{g} V_{fg} 
\nabla_{a}V^{de} \nabla^{a}V^{bc} + e^{3 \vp }  
 c_{18}  V_{d}{}^{f} V^{de} V_{e}{}^{g} V_{fg} 
\nabla^{a}V^{bc} \nabla_{b}V_{ac} \nn\\&&+ e^{3 \vp }  
 c_{21}  V_{c}{}^{e} V_{d}{}^{f} V_{e}{}^{g} V_{fg} 
\nabla^{a}V^{bc} \nabla_{b}V_{a}{}^{d} + e^{3 \vp }  
 c_{26}  V_{ad} V_{c}{}^{f} V_{e}{}^{g} V_{fg} 
\nabla^{a}V^{bc} \nabla_{b}V^{de} \nn\\&&+ e^{3 \vp }  
 c_{156}  V_{a}{}^{f} V_{c}{}^{g} V_{df} V_{eg} 
\nabla^{a}V^{bc} \nabla^{d}V_{b}{}^{e} + e^{3 \vp }  
 c_{157}  V_{a}{}^{f} V_{cf} V_{d}{}^{g} V_{eg} 
\nabla^{a}V^{bc} \nabla^{d}V_{b}{}^{e} \nn\\&&+ e^{3 \vp }  
 c_{158}  V_{ad} V_{c}{}^{f} V_{e}{}^{g} V_{fg} 
\nabla^{a}V^{bc} \nabla^{d}V_{b}{}^{e} + e^{3 \vp }  
 c_{180}  V_{ae} V_{bd} V_{c}{}^{g} V_{fg} \nabla^{a}V^{bc} 
\nabla^{d}V^{ef}\nn\\&& + e^{3 \vp }  
 c_{181}  V_{ad} V_{be} V_{c}{}^{g} V_{fg} \nabla^{a}V^{bc} 
\nabla^{d}V^{ef},
\eeqa
where the coupling constants \( c_1, \dots, c_{298} \) are determined in this work through the dimensional reduction of the relevant ten-dimensional gravity terms at order \( \alpha'^3 \). 

We determine the coefficients  by employing the scheme outlined in  \reef{T55}. This involves equating the result in \reef{ttee12} to our chosen minimal basis, once field redefinitions, integration by parts, and Bianchi identities have been applied. That is,
 \beqa
S_{IIA}&\sim&-\frac{2}{\kappa^2}\frac{\pi^3\alpha'^3}{2^{2}.3}\int d^{9}x\sqrt{-\bg}\,\cL(\vp,V)\,.\labell{ttee22}
\eeqa 
Here, $S_{IIA}$ represents the dimensionally reduced action from \reef{ttee12}, while $\mathcal{L}(\vp,V)$ denotes the specific minimal basis of 298 couplings identified in \reef{T55}. The $\sim$ relation signifies equality modulo:
\begin{itemize}
\item Field redefinitions
\item Total derivative terms
\item Bianchi identities
\end{itemize}
 From this matching, we uniquely determine all 298 coupling constants in \reef{T55}. These take the following values:
\beqa
&&c_{1} = 5471/432, c_{2} = -(11995/432), c_{3} = 9601/864, 
 c_{4} = 593/192, c_{5} = -(905/1728), \nn\\&&c_{6} = -(775/108), 
 c_{7} = 25543/1728, c_{8} = -(24031/13824), c_{9} = -(1903/1728), 
 c_{10} = 289/512, \nn\\&&c_{11} = 3235/432, c_{12} = 6227/864, 
 c_{13} = -(25543/864), c_{14} = 135935/3456, c_{15} = 157469/1728,\nn\\&& 
 c_{16} = 23963/1728, c_{17} = -(886/9), c_{18} = -(153385/13824), 
 c_{19} = 73385/3456, c_{20} = 3/64, \nn\\&&c_{21} = 5617/3456, 
 c_{22} = 235/54, c_{23} = 4271/864, c_{24} = -(2857/288), 
 c_{25} = 1/8, \nn\\&&c_{26} = -(871/432), c_{27} = -(4001/1728), 
 c_{28} = -(6119/432), c_{29} = 10699/864, \nn\\&&c_{30} = -(10375/864), 
 c_{31} = -(81/32), c_{32} = -(7/6), c_{33} = 4271/864, 
 c_{34} = -(7615/864), \nn\\&&c_{35} = 1369/768, c_{36} = -(58895/2304), 
 c_{37} = 5/4, c_{38} = 443/36, c_{39} = 31/96, c_{40} = 1/4, \nn\\&&
 c_{41} = -(3/8), c_{42} = -(7/16), c_{43} = -(1/8), c_{44} = -(5/4), 
 c_{45} = -(3/2), c_{46} = 5/4, \nn\\&&c_{47} = 10279/864, c_{48} = 1/4, 
 c_{49} = 1/2, c_{50} = 3/8, c_{51} = 3/8, c_{52} = 0, c_{53} = 5/16, 
 c_{54} = 3/4, \nn\\&&c_{55} = 0, c_{56} = 144737/6912, c_{57} = 719/256, 
 c_{58} = -(48845/864), c_{59} = -(49129/6912), \nn\\&&c_{60} = 75221/1728, 
 c_{61} = -(9169/1152), c_{62} = -(74357/1728), c_{63} = 9601/576, \nn\\&&
 c_{64} = -(39961/1728), c_{65} = -(1763/18), c_{66} = -(61/32), 
 c_{67} = 2551/1728, \nn\\&&c_{68} = -(6011/864), c_{69} = 52415/3456, 
 c_{70} = -(2443/1728), c_{71} = -(10847/6912), \nn\\&&c_{72} = 74573/1728, 
 c_{73} = 10465/2304, c_{74} = -(1763/36), c_{75} = 547/32, 
 c_{76} = 7079/1152,\nn\\&& c_{77} = -(75005/1728), c_{78} = -(9601/2304), 
 c_{79} = -(6443/1728), c_{80} = 1763/18, \nn\\&&c_{81} = -(363905/10368), 
 c_{82} = 24233/1728, c_{83} = -(359/128), c_{84} = 25111/1728,\nn\\&& 
 c_{85} = -(320/27), c_{86} = -(33269/1728), c_{87} = 3535/576, 
 c_{88} = -(25111/6912),\nn\\&& c_{89} = 5579/6912, c_{90} = 7/4, 
 c_{91} = -(5147/3456), c_{92} = 689/216, c_{93} = -(12191/864),\nn\\&& 
 c_{94} = -(11851/144), c_{95} = -(30659/3456), c_{96} = -(7/16), 
 c_{97} = -(141725/6912),\nn\\&& c_{98} = -(30263/4608), c_{99} = 239/54, 
 c_{100} = 39101/13824, c_{101} = 28373/2304, \nn\\&&
 c_{102} = -(109163/6912), c_{103} = 13831/1728, 
 c_{104} = -(304643/20736), c_{105} = 171791/6912, \nn\\&&
 c_{106} = 145241/10368, c_{107} = 44215/3456, c_{108} = 585167/41472,
  c_{109} = 11563/864, \nn\\&&c_{110} = 616085/10368, c_{111} = 7603/4608, 
 c_{112} = -(3535/72), c_{113} = 13341/128, \nn\\&&c_{114} = 48427/1728, 
 c_{115} = -(359273/13824), c_{116} = 3773/432, c_{117} = 1/2, \nn\\&&
 c_{118} = 10429/1728, c_{119} = -(11725/1728), c_{120} = 157901/864, 
 c_{121} = 886/9, \nn\\&&c_{122} = -(1763/18), c_{123} = 21371/1728, 
 c_{124} = 13021/3456, c_{125} = -(443/9), \nn\\&&c_{126} = 13703/864, 
 c_{127} = -(14675/4608), c_{128} = 9527/1296, c_{129} = -(12601/192),\nn\\&&
  c_{130} = 25147/576, c_{131} = 27581/1728, c_{132} = -(13489/1152), 
 c_{133} = 25147/1152, \nn\\&&c_{134} = 95555/1728, c_{135} = -(158261/1152),
  c_{136} = 8351/432, c_{137} = 26987/3456, \nn\\&&c_{138} = -(22883/3456), 
 c_{139} = 0, c_{140} = 12067/576, c_{141} = -(263/72), 
 c_{142} = -(6541/864), \nn\\&&c_{143} = -(253033/2304), 
 c_{144} = -(5579/864), c_{145} = -(271237/1152), \nn\\&&
 c_{146} = -(531631/2304), c_{147} = 271309/2304, 
 c_{148} = 30541/1728, c_{149} = 1523/54, \nn\\&&c_{150} = -(132445/10368), 
 c_{151} = -(2344/81), c_{152} = -(2245/432), c_{153} = 40933/3456, \nn\\&&
 c_{154} = 13129/1728, c_{155} = -(267649/3456), c_{156} = -(307/108),
  c_{157} = 749/216, \nn\\&&c_{158} = 997/1152, c_{159} = -(503/288), 
 c_{160} = 8539/3456, c_{161} = -(5243/432), \nn\\&&c_{162} = -(563/288), 
 c_{163} = 1/4, c_{164} = 11563/864, c_{165} = 4595/432, 
 c_{166} = -(10699/864), \nn\\&&c_{167} = 4/3, c_{168} = -(4595/864), 
 c_{169} = 10699/1728, c_{170} = 1349/216, c_{171} = 5887/864, \nn\\&&
 c_{172} = -(12319/432), c_{173} = 1/2, c_{174} = -(113/18), 
 c_{175} = 7687/864, c_{176} = -(11887/432), \nn\\&&c_{177} = -(5/16), 
 c_{178} = 95/18, c_{179} = -(12211/1152), c_{180} = -(811/864), \nn\\&&
 c_{181} = -(1313/3456), c_{182} = 2251/864, c_{183} = -(2467/864), 
 c_{184} = -(2467/864), c_{185} = 1/8, \nn\\&&c_{186} = 133/27, 
 c_{187} = -(4379/864), c_{188} = 293/54, c_{189} = -(17/24), 
 c_{190} = -(4937/432), \nn\\&&c_{191} = -(266/27), c_{192} = 142/9, 
 c_{193} = -(775/108), c_{194} = 22883/3456,  \nn\\&&c_{195} = 136567/10368,
 c_{196} = -(300047/55296), c_{197} = 6211/3456,  \nn\\&&
 c_{198} = -(3952841/165888), c_{199} = -(16099/3456), c_{200} = 1/8, 
 c_{201} = 1733959/20736,  \nn\\&&c_{202} = -(133/27), c_{203} = 251545/10368,
  c_{204} = 2037649/41472, c_{205} = 1/16, \nn\\&& c_{206} = -(158423/9216), 
 c_{207} = 1/32, c_{208} = 3/16, c_{209} = 1159/54, c_{210} = 0,  \nn\\&&
 c_{211} = 19609/6912, c_{212} = 11275/1152, c_{213} = -(5579/864), 
 c_{214} = 82007/13824,  \nn\\&&c_{215} = -(25165/1728), c_{216} = -(1/8), 
 c_{217} = 103321/6912, c_{218} = -(775/108),  \nn\\&&c_{219} = -(8875/1728), 
 c_{220} = 401/54, c_{221} = -(23/64), c_{222} = -(12211/864),  \nn\\&&
 c_{223} = 5209/108, c_{224} = -(12319/432), c_{225} = 3055/144, 
 c_{226} = -(14717/864), \nn\\&& c_{227} = 12427/864, c_{228} = 24233/864, 
 c_{229} = 11833/432, c_{230} = 11923/1152, c_{231} = 15/32,  \nn\\&&
 c_{232} = -(1/2), c_{233} = 22883/3456, c_{234} = 4697/288, 
 c_{235} = -(13/16), c_{236} = -(22883/3456),  \nn\\&&c_{237} = -(11815/576), 
 c_{238} = -(1/16), c_{239} = -(11923/288), c_{240} = 11923/576, 
 c_{241} = 11/8,  \nn\\&&c_{242} = 11635/576, c_{243} = 11563/1728, 
 c_{244} = 9997/1728, c_{245} = -(3635/288),  \nn\\&&c_{246} = 7795/1728, 
 c_{247} = 721/108, c_{248} = -(11995/432), c_{249} = 0, 
 c_{250} = 1/8, c_{251} = 0,  \nn\\&&c_{252} = -(7/64), c_{253} = 0, 
 c_{254} = -1, c_{255} = -1, c_{256} = 1/32, c_{257} = 141/128, 
 c_{258} = 149/576,  \nn\\&&c_{259} = -(515/108), c_{260} = 10267/1728, 
 c_{261} = -(2933/1728), c_{262} = -(559/54),  \nn\\&&c_{263} = 239/216, 
 c_{264} = -(11563/864), c_{265} = 293/108, c_{266} = 1/16, 
 c_{267} = 1/4, c_{268} = 0,  \nn\\&&c_{269} = -(1891/6912), 
 c_{270} = -(9403/864), c_{271} = 10699/864, c_{272} = 517/4608,  \nn\\&&
 c_{273} = -(2269/6912), c_{274} = 7361/864, c_{275} = -(251/576), 
 c_{276} = -(143/576),  \nn\\&&c_{277} = -(133/108), c_{278} = -(1/4), 
 c_{279} = 5/8, c_{280} = -(11293/1728), c_{281} = -(3851/3456),  \nn\\&&
 c_{282} = 0, c_{283} = 0, c_{284} = 37001/6912, c_{285} = 17/32, 
 c_{286} = 2339/13824, c_{287} = -(1/4),  \nn\\&&c_{288} = 16339/1152, 
 c_{289} = -(803/216), c_{290} = -(16051/1728), c_{291} = 25465/6912, \nn\\&& 
 c_{292} = 175/1728, c_{293} = 937/54, c_{294} = -(3631/1728), 
 c_{295} = -(3181/108),  \nn\\&&c_{296} = 6497/144, c_{297} = 25543/864, 
 c_{298} = -(3427/288)\labell{sol}
\eeqa
    Of the 298 couplings in the minimal basis, 288 are non-zero and 10 are zero. Among the vanishing coefficients is \( c_{268} \), which is the sole independent coupling involving the Ricci tensor.


\end{document}